\documentclass[aps,prb,floatfix,amsmath,amssymb,preprint,superscriptaddress,eqsecnum]{revtex4}
\usepackage[dvips]{graphicx}
\usepackage{dcolumn}
\usepackage{bm}
\usepackage{setspace}   

\begin{document}
\title{Hole dynamics in an antiferromagnet\\ across a deconfined quantum critical point}

\author{Ribhu K. Kaul}
\affiliation{Department of Physics, Harvard University, Cambridge MA
02138, USA}

\author{Alexei Kolezhuk}
\affiliation{Department of Physics, Harvard University, Cambridge MA
02138, USA}
\affiliation{Institut f\"ur Theoretische Physik,
Universit\"at Hannover, 30167 Hannover, Germany}

\author{Michael Levin}
\affiliation{Department of Physics, Harvard University, Cambridge MA
02138, USA}

\author{Subir Sachdev}
\affiliation{Department of Physics, Harvard University, Cambridge MA
02138, USA}

\author{T. Senthil}
\affiliation{Center for Condensed Matter Theory, Department of
Physics, Indian Institute of Science, Bangalore 560 012, India}
\affiliation{Department of Physics, Massachusetts Institute of
Technology, Cambridge MA 02139, USA}

\date{January 2007\\[24pt]}

\begin{abstract}
We study the effects of a small density of holes, $\delta$, on a
square lattice antiferromagnet undergoing a continuous transition
from a N\'eel state to a valence bond solid at a deconfined quantum
critical point. We argue that at non-zero $\delta$, it is likely
that the critical point broadens into a non-Fermi liquid `holon
metal' phase with fractionalized excitations. The holon metal phase
is flanked on both sides by Fermi liquid states with Fermi surfaces
enclosing the usual Luttinger area. However the electronic
quasiparticles carry distinct quantum numbers in the two Fermi
liquid phases, and consequently the ratio $\lim_{\delta \rightarrow
0} \mathcal{A}_F/\delta$ (where $\mathcal{A}_F$ is the area of a
hole pocket) has a factor of 2 discontinuity across the quantum
critical point of the insulator. We demonstrate that the electronic
spectrum at this transition is described by the `boundary' critical
theory of an impurity coupled to a 2+1 dimensional conformal field
theory. We compute the finite temperature quantum-critical
electronic spectra and show that they resemble ``Fermi arc'' spectra
seen in recent photoemission experiments on the pseudogap phase of
the cuprates.
\end{abstract}




%

\maketitle

\section{Introduction}
\label{sec:intro}

There was a great deal of work on the dynamics of a single hole in a
square lattice antiferromagnet, soon after the discovery of high
temperature superconductivity in the cuprate compounds. It was
demonstrated\cite{trugman,ssneel,elser,klr,martinez,poilblanc,reiter,mish}
that a single hole moving in a N\'eel ground state has a finite
quasiparticle residue, $Z$; so a small density of holes, $\delta$,
are expected to form a Fermi liquid. This Fermi liquid state with
N\'eel order will be the starting point of our analysis. Also,
Shraiman and Siggia \cite{ss,wiese} introducing a current-current
coupling between the hole and the antiferromagnet which implied that
a large spin $S$ N\'eel state is unstable for certain parameter
ranges to spiral spin ordering; we shall {\em not\/} be interested
in this metallic spiral state here, although the Shraiman-Siggia
coupling (in Eq.~(\ref{lc1}) below) will play a key role in our
analysis.

We begin with a $S=1/2$ N\'eel state of an insulating
antiferromagnet and imagine ``turning up quantum fluctuations'' by
adding further neighbor or ring-exchange couplings so that there is
a transition to a paramagnetic state in which spin rotation
invariance is restored. Now add a small density of holes to this
antiferromagnet. The main question we shall address is: what is the
fate of the Fermi liquid N\'eel state across such a transition\,?

\begin{figure}
\centering
  \includegraphics[width=5.5in]{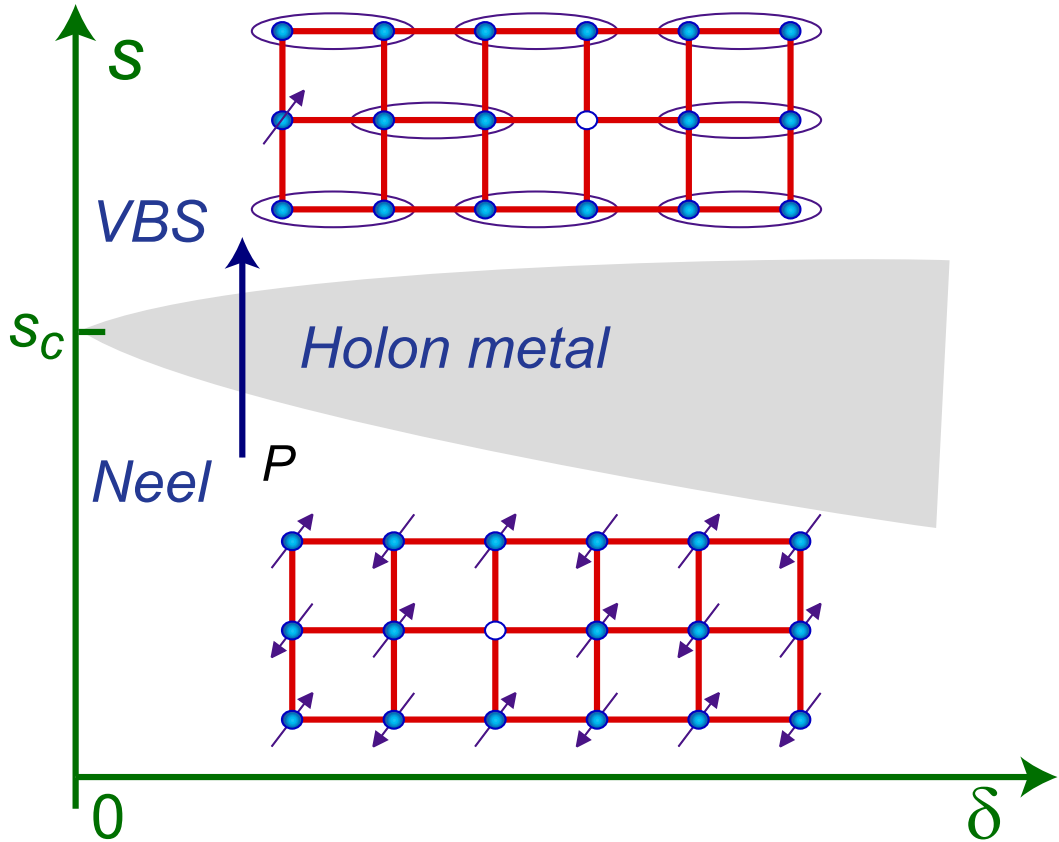}
  \caption{Schematic phase diagram. The ellipses represent spin singlet valence
  bonds. The coupling $s$ tunes the insulator across the N\'eel-VBS
  transition, and $\delta$ is the mobile hole density.
  The deconfined quantum critical point is at $s=s_c$ in the
  insulator with $\delta=0$. The vacancies (`holons') carry a gauge charge $q=\pm 1$
  under an emergent U(1) gauge force. In the cartoons above,
  the reader can interpret $q$ as a sublattice label.
  In the $s < s_c$, N\'eel phase,
  $q$ determines the spin: a vacancy on an up (down) spin site
  carries net spin down (up) and so is equivalent to a charge $e$ spin-1/2 hole.
  For $s>s_c$, the hole is a composite of a vacancy and
  a nearby unpaired spin with opposite $q$, moving by rearranging
  nearest-neighbor valence bonds; note that this motion preserves spin and sublattice quantum
  numbers {\em separately\/} (see also Ref.~\onlinecite{ls}). So
  there are twice
  as many states per momentum for a charge $e$ spin 1/2 hole
  in the VBS state than there are in the N\'eel state.}
  \label{fig1}
\end{figure}

Specifically, we consider the `deconfined' quantum phase transition
proposed in Ref.~\onlinecite{senthil1} for an insulating $S=1/2$
square lattice antiferromagnet\cite{proko,anders,kamal,mv}. This is
a theory for a transition between a N\'eel state and a spin-gap
state with valence bond solid (VBS) order (the latter state is spin
rotation invariant, but breaks lattice symmetries by ordering of
valence bonds). These two states break distinct symmetries of the
Hamiltonian, and so cannot generically have a continuous transition
between them in the Landau-Ginzburg-Wilson theory of phase
transitions. However, such a transition is found in a `deconfined'
theory focusing not on order parameters but on fractionalized
excitations and emergent gauge forces. The transition is tuned by
the coupling $s$ (which represents the strength of frustrating
exchange interactions)---see Fig.~\ref{fig1}.
Upon doping, we will argue that the most likely possibility is that
the insulating deconfined critical point gets broadened into a novel
non-Fermi liquid `holon metal' phase, with no Fermi surface (shown
shaded in Fig.~\ref{fig1}). The qualitative distinction between the
N\'eel and VBS states survives also in the Fermi liquid states at
$\delta >0$ (shown unshaded in Fig.~\ref{fig1}): we will show that a
characteristic property (specified shortly) of the Fermi surface has
a discontinuity in the limit $\delta \rightarrow 0$ as $s$ is
scanned across the $s=s_c$ critical point of the insulator. We also
compute finite temperature electronic spectra in the vicinity of the
transition and find that they resemble ``Fermi arc'' spectra seen in
recent photoemission experiments on the pseudogap phase of the
cuprates\cite{mohit}.

There have been other discussions in the
literature\cite{coleman0,qsi0,senthil0} of transitions between Fermi
liquids, including proposals that there could be a continuous
quantum transition with a discontinuous change in the shape of the
Fermi surface (recent experiments \cite{onuki} on CeRhIn$_5$ are
compatible with an abrupt or very rapid change in Fermi surface
topology). This would require a sudden change in the Fermi surface
as a function of $s$ at a fixed non-zero value of $\delta$. We will
argue that such a change is unlikely in our models, and the
situation is as illustrated in Fig.~\ref{fig1}, with an intermediate
non-Fermi liquid phase.

By an extension of arguments in early work
\cite{wensc,palee,shankar,sarker}, it is expected that a significant
portion of the phase diagram in Fig.~\ref{fig1} is unstable at low
temperatures to superconductivity. We defer consideration of such
superconducting states to future work, and limit ourselves here to
the normal states.

We now turn to a more detailed summary of our results. First, we
discuss our results in the unshaded regions of Fig~\ref{fig1}. In
these regions, we are adding a small density of mobile carriers to
conventionally ordered insulators, and we obtain Fermi liquid phases
with electron-like quasiparticles with a non-zero quasiparticle
residue, {\em i.e.\/} $Z \neq 0$. Non-Fermi-liquid physics appears
only in the shaded region.

\begin{figure}
\centering
  \includegraphics[width=5.5in]{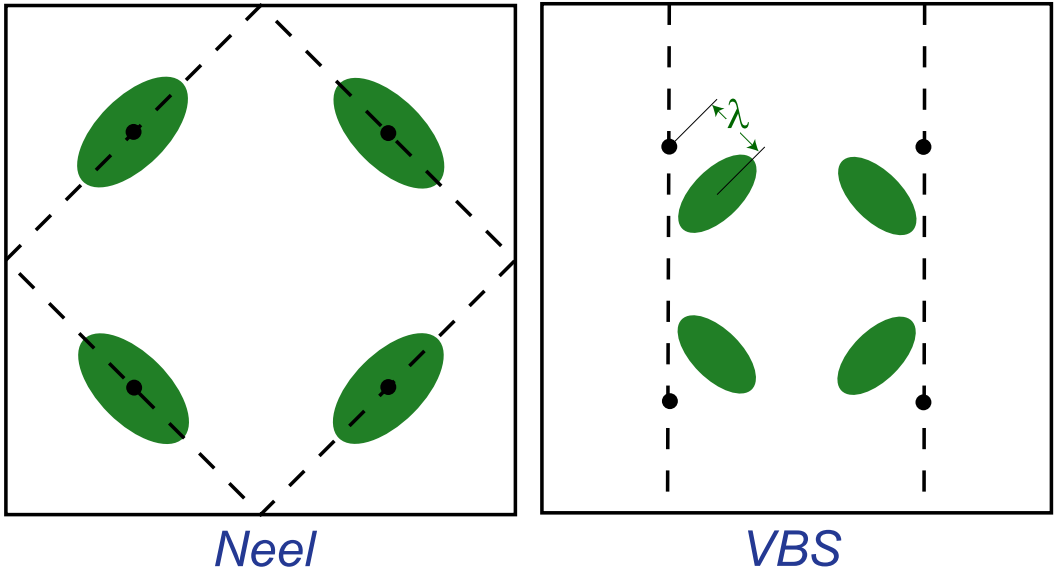}
  \caption{Momentum space Fermi surfaces in the N\'eel and VBS regions of
  Fig~\ref{fig1}. The filled circles are the 4 $\vec{K}_p$
  wavevectors, with $\vec{K}_1 = (\pi/2a) (1,1)$, $\vec{K}_2 =
  (\pi/2a) (1,-1)$, $\vec{K}_3 = - \vec{K}_1$, $\vec{K}_4 = -
  \vec{K}_2$, with $a$ the lattice spacing.
  The dashed line in the N\'eel phase indicates the boundary of the
  magnetic Brillouin zone. Only the Fermi surfaces within this zone
  contribute to the Luttinger counting, and so the area of each
  ellipse is $\mathcal{A}_F = (2 \pi)^2 \delta/4$. In the VBS phase,
  all 4 pockets are inequivalent, and so the area of each ellipse is
  $\mathcal{A}_F = (2 \pi)^2 \delta/8$. The dashed lines now show
  the reduction of the Brillouin zone due to the VBS order which appears at sufficiently
  low temperatures;
  ``shadow'' Fermi sufaces, with weak photoemission intensity
  (estimated in the text), will appear as
  reflections across these lines, and these Fermi surfaces are not
  shown.
  }
  \label{fig2}
\end{figure}

In the $s<s_c$ N\'eel phase, we obtain a Fermi liquid
state\cite{trugman,ssneel,elser,klr,martinez,poilblanc,reiter,mish}
with four hole pockets centered at the $\vec{K}_p = (\pi/2a) (\pm 1,
\pm 1)$, where $a$ is the lattice spacing, shown in Fig.~\ref{fig2}.
However, because of the halving of the Brillouin zone by magnetic
order, only two of these pockets are distinct. After accounting for
the two-fold spin degeneracy, we conclude that the area enclosed by
each pocket is $\mathcal{A}_F = (2 \pi)^2 \delta/4$. Another way of
understanding this halving of the Brillouin zone (which is also
indicated in the caption of Fig.~\ref{fig1}, and discussed further
in Section~\ref{sec:neel}) is as follows. We can consider the doped
hole as a vacancy in the background of a N\'eel state. If this hole
is to move without leaving a string of broken bonds\cite{trugman},
it must preserve its sublattice label. However, because of the
broken symmetry associated with the N\'eel order, the sublattice
location is not independent of the spin of the vacancy, and two
labels are really the same quantum number.

Next, we discuss a small density of holes in the $s>s_c$ VBS state.
As we will demonstrate in Section~\ref{sec:vbs}, in this state the
four hole pockets are no longer pinned at the $\vec{K}_p$, but
instead move a distance $\lambda$ away, as indicated in
Fig.~\ref{fig2}. This shift arises from the Shraiman-Siggia
\cite{ss} coupling. The value of $\lambda$ is determined by $s-s_c$,
but is {\em independent\/} of $\delta$ to lowest order in $\delta$.
Consequently, for sufficiently small $\delta$, the hole pockets do
not intersect the reduced Brillouin zone boundaries, associated with
the appearance of VBS order. The four hole pockets therefore all
contain distinct quasiparticles states, and after accounting for the
two-fold spin degeneracy, we now conclude that the area enclosed by
each pocket is $\mathcal{A}_F = (2 \pi)^2 \delta/8$. As above,
another interpretation of this result is indicated in
Fig~\ref{fig1}, and will be described in more detail in
Section~\ref{sec:vbs}: the hole motion in the VBS state also
preserves its sublattice index, but now the sublattice and spin
labels are distinct quantum numbers.

We can summarize the above statements into one of the main zero
temperature results of this paper:
\begin{equation}
\lim_{\delta \rightarrow 0}  \left. \frac{\mathcal{A}_F}{\delta}
\right|_{s<s_c} = 2 \times \lim_{\delta \rightarrow 0} \left.
\frac{\mathcal{A}_F}{\delta} \right|_{s>s_c} \label{ratio}
\end{equation}
Note that on both sides of the equation, we are taking the limit
$\delta \rightarrow 0$ at {\em fixed\/} $s$. Thus, although a
characteristic feature of the Fermi surface (the ratio
$\mathcal{A}_F /\delta$) changes discontinuously as $s$ crosses
$s_c$, the Fermi surface itself is of vanishingly small size. The
result in Eq.~(\ref{ratio}) does not constitute a discontinuous
change in the Fermi surface in the sense of other
proposals\cite{coleman0,qsi0,senthil0}.

Next, we address the evolution of the Fermi surface along the {\em
fixed\/} $\delta>0$ line $P$ in Fig~\ref{fig1}. Aspects of the
physics will be addressed in Section~\ref{sec:qc}, and some
questions are deferred to future work. The key issue is the fate of
the monopoles in the U(1) gauge theory which describes the
deconfined critical point of the insulator at $s=s_c$. The situation
has been discussed at length elsewhere \cite{senthil1} for
$\delta=0$: the monopoles are irrelevant at the $s=s_c$ critical
point, but are relevant for all $s>s_c$ where they induce
confinement and VBS order. There are two distinct interesting
possibilities for the fate of this transition for
$\delta > 0$ (a first-order transitions is, of course, also possible):\\
({\em i\/}) The critical point at $s=s_c$, $\delta=0$ extends into a
single line of deconfined critical points for $\delta= 0$, with
Fermi-liquid physics on either side of the line. In this case,
physics discussed above Eq.~(\ref{ratio}) will apply also for
$\delta > 0$, and there will be a discontinuous Fermi surface change
across a continuous transition, realizing scenarios postulated in
previous work \cite{coleman0,qsi0,senthil0}. However, we will argue
in Section~\ref{sec:qc} that this possibility
is unlikely in the present model.\\
({\em ii\/}) The deconfined critical point at $s=s_c$, $\delta=0$
broadens into a deconfined phase, extending over a finite range of
$s$ values (shown as the shaded region in Fig.~\ref{fig1}), with
novel non-Fermi-liquid physics. The monopoles, which were suppressed
by the gapless critical modes at the single point $s=s_c$ in the
$\delta=0$ insulator, are now suppressed \cite{hermele} over a
finite range of values of $s$ by the additional gapless excitations
associated with mobile charge carriers for $\delta > 0$. Arguments
supporting this possibility will appear in Section~\ref{sec:qc}. For
reasons which will become clear in Section~\ref{sec:qc}, we will
refer to this phase as a `holon metal'. This holon metal phase has
no electron-like Fermi surfaces, and hence $Z=0$.


It is convenient to think about the phase diagram in the $s$-$\mu$
plane where $\mu$ is the chemical potential. For a range of $\mu$ in
the Mott insulator the ground state does not change for fixed $s$.
The point $s = s_c$ in this Mott insulator is the deconfined quantum
critical point and its low energy theory is a conformal field theory
(CFT). Below some critical $\mu_c$ the hole density $\delta$ holes
will increase from zero as some power of $\mu - \mu_c$. It should be
clear that the critical point at $s = s_c$, $\mu = \mu_c$ is an
unstable  multicritical point which controls the properties of the
holon metal phase at low doping. In this perspective\cite{vojta}, we
view $\mu_c - \mu$ as a relevant perturbation to the field theory
describing this multicritical point. So at an energy, temperatures,
or wavevectors higher than a low energy scale determined by
$\delta$, the physics is described by the response of the CFT of the
deconfined critical point of the Mott insulator to an infinitesimal
hole density. We will argue in Section~\ref{sec:qc} that this
response is associated with a `boundary CFT' describing a quantum
impurity in the bulk, 2+1 dimensional CFT. The result in
Eq.~(\ref{ratio}) describing the Fermi liquid phases outside the
shaded region of Fig.~\ref{fig1} can also be viewed as a consequence
of the structure of this boundary CFT.

Section~\ref{sec:photo} will explore practical consequences of the
boundary CFT by presenting numerical results on the form of the
finite temperature electronic spectral weight. This will be carried
out using the full action presented in Section~\ref{sec:ft},
including terms that are formally corrections to scaling to the
boundary CFT, but are important for experimental comparisons. We
will find that the results resemble recent photoemission
measurements on the pseudogap phase of the cuprates\cite{mohit}; in
particular, the length of the ``Fermi arcs'' is roughly proportional
to the temperature. Furthermore, we find that the
Shraiman-Siggia\cite{ss} term, which was responsible for the shift
in the centers of the hole pockets away from the $\vec{K}_p$ in the
zero temperature VBS state (Section~\ref{sec:vbs} and
Fig.~\ref{fig2}), also shifts the centers of the finite temperature
``Fermi arcs'' from the $\vec{K}_p$, as is seen experimentally. This
feature is not present in a recent theory based upon a
`staggered-flux' spin liquid \cite{arun}.

We will begin in Section~\ref{sec:ft} by presenting the field theory
which describes the CFT at $s=s_c$, $\delta=0$ and the perturbations
which induce mobile carriers for $\delta > 0$. The Fermi liquid
phases of this field theory will be discussed in
Sections~\ref{sec:neel} and~\ref{sec:vbs} for the N\'eel and VBS
states respectively. The non-Fermi-liquid holon metal phase, and the
finite temperature quantum criticality is described in
Section~\ref{sec:qc}, and we conclude in Section~\ref{sec:conc}.

\section{Field theory}
\label{sec:ft}

There is a vast literature on the theory of the lightly doped
antiferromagnet. Two main approaches have been taken. The
`spin-fermion' models \cite{chubukov}, which are extrapolations of
the weak-coupling Hartree-Fock theory of the Hubbard model, offer a
convenient description of the Fermi liquid states with N\'eel order.
However, these models cannot describe the transition to the VBS
state, and so are not suitable for our purposes. The second
approach, \cite{ss,wiese} departs from the spin-wave theory of the
insulating antiferromagnet, and describes the coupling of the holes
to a spin deformations of the antiferromagnet. In modern terms, this
is formulated as `chiral perturbation theory' in which the broken
symmetry of the N\'eel state allows one to place constraints on the
couplings between the mobile holes and the `chiral' spin wave
excitations\cite{wiese}. However, the non-linear realization of the
SU(2) spin rotation symmetry in the N\'eel state makes this approach
cumbersome. For our purposes, a central shortcoming is that this
approach is not able to describe the fermion spectrum in a state in
which SU(2) symmetry is restored and the magnetic Brillouin zone
expands to the full Brillouin zone of the square lattice: Bloch
reflections across the magnetic Brillouin zone boundary do not
disappear at any order in chiral perturbation theory. It is
essential to considerations of Fermi surface topology that such
effects are properly accounted for.

We will instead use a method in which the spin SU(2) symmetry is
linearly realized and is kept explicit at all stages. We also want
to explicitly preserve the full space group symmetry of the square
lattice. In the insulator, this requires starting from a `spin
liquid' state. Although there is no gapped spin liquid insulator in
Fig~\ref{fig1}, the VBS state descends \cite{rs,senthil1} from an
instability of a particular U(1) spin liquid, \cite{aa} and the
latter will serve as our starting point. Also, the CFT at $s=s_c$,
$\delta=0$, defines an `algebraic spin liquid', and this is obtained
as a gapless critical point in the theory of the same U(1) spin
liquid.

It has been emphasized by Wen \cite{wenpsg} that an essential
characteristic of any spin liquid is its PSG: the projective
symmetries of the various fields under elements of the square
lattice space group. Because the fields carry charges under a gauge
group which characterizes the spin liquid, they need not be
invariant under such transformations---they are only required to be
invariant up to a gauge transformation, and this is sufficient to
preserve square lattice symmetry in all observables. Here we will
show how the PSG can be extended from the insulator to the doped
state, where it also places crucial constraints on the low energy
effective theory. Another PSG-based analysis of a doped Mott
insulator is in Ref.~\onlinecite{vortexpsg} for case where the spin
liquid has fermionic spinons. Here we focus on the vicinity of the
N\'eel-VBS transition where, as we review below, the spinons are
bosonic.

The motion of charge carriers doped into an insulating
square-lattice quantum antiferromagnet is conventionally described
by the ``$t$-$J$'' model,
\begin{equation}
\label{tJmodel} H_{t-J} = -
\sum_{i,j,\alpha}t_{ij}(c^{\alpha\dagger}_{i}c_{j\alpha} + {\rm
h.c.}) + \sum_{i,j} J_{ij}\vec{S}_i \cdot \vec{S}_j + \ldots \, ,
\end{equation}
where $c^{\alpha\dagger}_{i}$ creates an electron with spin $\alpha$
on the sites $i$ of a square lattice and
$\vec{S}_i=\frac{1}{2}\sum_{\alpha\beta} c^{\alpha\dagger}_{i}
\vec{\sigma}_{\alpha}^{\beta}c_{i\beta}$, with $\vec{\sigma}$ the
Pauli matrices. In addition, the constraint
$\sum_{\alpha}c^{\alpha\dagger}_{i}c_{i\alpha}\leq 1$ is enforced on
each site, modeling the large local repulsion between the electrons.
It is important to note that our results are more general than a
particular $t$-$J$ model, and follow almost completely from symmetry
considerations. The ellipses in Eq.~(\ref{tJmodel}) additional
short-range couplings which preserve square lattice symmetry and
spin rotation invariance. The coupling $s$ is some axis in this
multidimensional parameter space which accesses both the N\'eel and
spin liquid phases in the insulator.

We will analyze $H_{t-J}$ by choosing a representation of the
electron operator which obeys the constraint on each site, and most
conveniently accesses the U(1) spin liquid describing the N\'eel-VBS
transition. This is achieved by introducing neutral Schwinger bosons
to represent the spin degrees of freedom, and spinless fermionic
holons which track the charge. Although this is not essential, we
will use distinct representations on the two square sublattices
because this simplifies the mean-field structure of the spin liquid
solution. On one sublattice of the square lattice we write the
electron operator, $c_\alpha$ as
\begin{equation}
c_\alpha = b_\alpha f^\dagger_+ \label{e1aa}
\end{equation}
where $b_\alpha$ are canonical Schwinger bosons and $f_+$ are
canonical fermionic holons. The constraint on each site now becomes
\begin{equation}
f_+^\dagger f_+ +  b^{\alpha\dagger} b_{ \alpha} = 1 \, .\label{e1}
\end{equation}
On the other sublattice, we use bosons which transform as a
conjugate respresentation
\begin{equation}
c_\alpha = \varepsilon_{\alpha\beta} \overline{b}^\beta f^\dagger_-
\label{e1a}
\end{equation}
with a similar constraint. We define the antisymmetric tensor
$\varepsilon_{\alpha\beta}$ by $\varepsilon_{\uparrow\downarrow}=1$
and $\varepsilon_{\alpha\beta}=\varepsilon^{\alpha\beta}$. Notice
that the representations in Eq.~(\ref{e1aa},\ref{e1a}) have a U(1)
gauge redundancy. The structure of the mean-field theory described
below instructs us to take the continuum limit so that the U(1)
gauge charges of $b_\alpha$ and $f_+$ are $+1$, while those of the
$\overline{b}^\alpha$ and $f_{-}$ are $-1$.

First, we recall the mean field theory \cite{aa} of $H_{t-J}$ in the
insulator in the above representation. Here we can ignore the holons
$f_{\pm}$. The exchange interactions are quartic in the Schwinger
boson operators, and we decouple these by a Hubbard-Stratanovich
transformation. This yields the mean field Hamiltonian for the U(1)
spin liquid
\begin{equation}
\label{eq:schwinger} H_J =  - Q \sum_{\langle ij\rangle} b_{i\alpha}
\overline{b}_{j}^{\alpha} + \mbox{H.c.} + \lambda \sum_{i}
b_i^{\alpha\dagger} b_{i \alpha} + \lambda \sum_{j}
\overline{b}_{j\alpha}^{\dagger} \overline{b}_{j}^{\alpha},
\end{equation}
where $i$ is restricted to be on one sublattice, and $j$ on the
other, and $Q$, $\lambda$ are positive constants to be determined by
solving the mean-field equations.  Upon adding frustrating exchange
couplings to $H_J$, the mean field parameters will vary, allowing us
to access both the N\'eel and spin liquid phases. For sufficiently
large $\lambda$, the boson spectrum is gapped, and this describes
the spin liquid ($s>s_c$). As $\lambda$ is decreased, the bosons
eventually condense, leading to a transition to the N\'eel state. We
now want to obtain a continuum field-theoretic representation of
this quantum phase transition of the insulator. This can done by
carefully taking the continuum limit of the low energy excitations
of $H_J$, and then including fluctuations about the mean-field
saddle point: such a procedure is described in detail in
Ref.~\onlinecite{rs}. Here, we show that the same answer can be
rapidly obtained by a PSG analysis. For this, we need to understand
how $H_J$ preserves the full square lattice symmetry even though it
treats the two sublattices inequivalently. So we consider the
transformations of the operators under all the space group
operations of the square lattice, and also under time-reversal.
These are listed in Table~\ref{table0}.
\begin{table}[t]
\begin{spacing}{2}
\centering
\begin{tabular}{||c||c|c|c|c||} \hline\hline
 & $T_x$ & $R_{\pi/2}^{\rm dual}$ & $I_x^{\rm dual}$ & $\mathcal{T}$  \\
 \hline\hline
$~~b_\alpha$~~ & $ ~\varepsilon_{\alpha\beta} \overline{b}^{\beta}~$
& $ ~\varepsilon_{\alpha\beta} \overline{b}^{\beta }~$ & $
~\varepsilon_{\alpha\beta} \overline{b}^{\beta }~$ & $
~\varepsilon_{\alpha\beta} b^{\beta \dagger}~$
\\
\hline $\overline{b}^\alpha$ & $ \varepsilon^{\alpha\beta}
{b}_{\beta}$ & $ \varepsilon^{\alpha\beta} {b}_{\beta }$ & $
\varepsilon^{\alpha\beta} {b}_{\beta }$ & $
\varepsilon^{\alpha\beta} \overline{b}_{\beta}^{\dagger}$
\\
\hline
$f_{+}$ & $ f_{-}$ & $ f_{-}$ & $ f_{-}$  &  $f_{+}^\dagger $ \\
\hline $f_{-}$ & $-f_{+}$ & $ -f_{+}$ & $ -f_{+}$  & $
f_{-}^\dagger$ \\ \hline\hline
\end{tabular}
\end{spacing}
\caption{PSG transformations of the lattice fields under square
lattice symmetry operations. $T_x$: translation by one lattice
spacing along the $x$ direction; $R_{\pi/2}^{\rm dual}$: 90$^\circ$
rotation about a dual lattice site on the plaquette center
($x\rightarrow y,y\rightarrow-x$); $I_x^{\rm dual}$: reflection
about the dual lattice $y$ axis ($x\rightarrow -x,y\rightarrow y$);
$\mathcal{T}$: time-reversal, defined as a symmetry (similar to
parity) of the imaginary time path integral. Note that such a
$\mathcal{T}$ operation is not anti-linear. The transformations of
the Hermitian conjugates are the conjugates of the above, except for
time-reversal of fermions \cite{vortexpsg}. For the latter,
$f_{\pm}$ and $f^\dagger_{\pm}$ are treated as independent Grassman
numbers and $\mathcal{T}: f^\dagger_{\pm} \rightarrow - f_{\pm}$.}
\label{table0}
\end{table}
 From the PSG of the Schwinger bosons, and using
Eqs.~(\ref{e1aa},\ref{e1a}), we can immediately deduce the PSG of
the lattice fermion fields $f_{\pm}$. These are also shown in
Table~\ref{table0}. Note especially the sign in the last row, which
is a consequence of
$\varepsilon^{\alpha\beta}\varepsilon_{\beta\gamma} =
-\delta^{\alpha}_{\gamma}$.

We now use the PSG to obtain the continuum theory of the N\'eel-VBS
transition in the insulator. The boson spectrum of $H_J$ shows that
the minimum energy excitations are near zero momentum, and so we may
take its low energy limit by a naive gradient expansion. We can
proceed in terms of the $b_\alpha$ and the $\overline{b}^\alpha$,
but it is convenient to introduce the linear combination
\begin{equation}
z_\alpha \sim b_\alpha + \overline{b}^{\dagger}_\alpha
\end{equation}
because the orthogonal linear combination can be integrated
out\cite{rs}. The PSG of the `spinons' $z_\alpha$ follows from
Table~\ref{table0} and is listed in Table~\ref{tablepsg}. Also
important for the low energy theory is a U(1) gauge field, $A_\mu$
($\mu=\tau,x,y$ is a spacetime index), which represents the
fluctuations of the phase of $Q$ and of $\lambda$. The $z_\alpha$
carries charge $+1$ under the $A_\mu$ gauge interaction. The
explicit PSG of the $A_\mu$ can be found elsewhere \cite{vortexpsg},
and we do not list it explicitly here because gauge-invariance is
sufficient to determine the $A_\mu$ terms.
\begin{table}[t]
\begin{spacing}{2}
\centering
\begin{tabular}{||c||c|c|c|c||} \hline\hline
 & $T_x$ & $R_{\pi/2}^{\rm dual}$ & $I_x^{\rm dual}$ & $\mathcal{T}$  \\
 \hline\hline
$z_\alpha$ & $ ~\varepsilon_{\alpha\beta} z^{\beta \ast}~$  & $
~\varepsilon_{\alpha\beta} z^{\beta \ast}~$ & $
~\varepsilon_{\alpha\beta} z^{\beta \ast}~$ & $
~\varepsilon_{\alpha\beta} z^{\beta \ast}~$
\\
\hline $f_{+1}$ & $i f_{-1}$ & $i f_{-2}$ & $i f_{-2}$  &  $f_{+1}^\dagger$ \\
\hline $f_{-1}$ & $-i f_{+1}$ & $i f_{+2}$ & $i f_{+2}$  & $ -f_{-1}^\dagger$ \\
\hline $f_{+2}$ & $i f_{-2}$ & $i f_{-1}$ & $i f_{-1}$  &  $f_{+2}^\dagger$ \\
\hline $f_{-2}$ & $-i f_{+2}$ & $-i f_{+1}$ & $i f_{+1}$  &  $- f_{-2}^\dagger$ \\
\hline $~\Phi_{\rm VBS}~$ & $-\Phi_{\rm VBS}^\ast $ & $i\Phi_{\rm
VBS}^\ast$ &
$\Phi_{\rm VBS}$ & $\Phi_{\rm VBS}$ \\
\hline\hline
\end{tabular}
\end{spacing}
\caption{PSG transformations of the fields in the continuum theory
under square lattice symmetry operations.    } \label{tablepsg}
\end{table}
Writing down the most general action invariant under the PSG we
obtain the theory of the low energy excitations of the spin liquid:
$\int d^2 r d\tau \mathcal{L}_J$ with
\begin{eqnarray}
\mathcal{L}_J &=& \mathcal{L}_z + \mathcal{L}_m \left[ \Phi_{\rm
VBS} \right]
\nonumber \\
 \mathcal{L}_z &=&|( \partial_\mu - i A_\mu ) z_\alpha
|^2 + s |z_\alpha |^2 + \frac{u}{2} \left( |z_\alpha|^2 \right)^2
\label{lz}
\end{eqnarray}
Here $s$ is the tuning parameter in Fig.~\ref{fig1}, we have set a
spinon velocity to unity, and $u$ is a spinon self-interaction. For
$s>s_c$, this field theory is in a spin SU(2)-invariant U(1) spin
liquid phase, with spinon excitations above an energy gap $\Delta$.
This gap vanishes as we approach the critical point as $\Delta \sim
(s-s_c)^\nu$, where $\nu$ is the correlation length exponent of the
CFT describing the deconfined critical point. For $s<s_c$, the
$z_\alpha$ condense into a Higgs phase, which is the N\'eel state of
the antiferromagnet. The N\'eel order parameter is $\vec{N} =
z^{\alpha \ast} \vec{\sigma}_\alpha^\beta z_\beta$, and this order
vanishes as $(s_c - s )^{\beta_{\rm Neel}}$.

The term $\mathcal{L}_m$ describes the dynamics of monopoles in the
compact U(1) gauge field $A_\mu$. This term has been discussed at
length elsewhere,\cite{rs,senthil1} and we will not reiterate the
details here. For our purposes, the important conclusions of these
earlier analyses are: ({\em i\/}) monopoles are suppressed for $s
\leq s_c$, in the N\'eel phase and at the deconfined quantum
critical point; ({\em ii\/}) monopoles proliferate in the $s>s_c$
spin liquid phase, inducing confinement of spinons and the
appearance of VBS order. A key fact \cite{rs,vortexpsg} which leads
to the latter conclusion is that Berry phases of the underlying
antiferromagnet endow the monopoles with a non-trivial
PSG\cite{haldane}. Indeed, this PSG allows us to
identify\cite{senthil1} the monopole creation operator with the VBS
order parameter, $\Phi_{\rm VBS}$. This order measures modulations
in the density of singlet valence bonds on the square lattice, and
is defined by
\begin{equation}
\Phi_{\rm VBS}(\vec{r})\equiv (-1)^{\vec{r}\cdot \hat{x}}
S_{\vec{r}} \cdot S_{\vec{r}+ \hat{x}} + i (-1)^{\vec{r}\cdot
\hat{y}} S_{\vec{r}} \cdot S_{\vec{r}+\hat{y}} \, , \label{defphi}
\end{equation}
where $\hat{x},\hat{y}$ are the unit lattice vectors. The
proliferation of monopoles implies that for $s>s_c$, $\langle
\Phi_{\rm VBS} \rangle \neq 0$. This order vanishes as we approach
the critical point,  $\langle \Phi_{\rm VBS} \rangle \sim
(s-s_c)^{\beta_{\rm VBS}}$, and the exponent is determined by the
scaling dimension of the monopole creation operator. The
representation in Eq.~(\ref{defphi}) also allows us to determine the
PSG of $\Phi_{\rm VBS}$, and hence of the monopoles, and this is
also listed in Table~\ref{tablepsg}.

Now let us move to the doped antiferromagnet, and examine the nature
of the low energy $f$ motion. The primary input we need from the
microscopic physics is the form of the dispersion of a single $f$
fermion in the Brillouin zone in the background of the U(1) spin
liquid just described. For $t \gg J$, this leads to a
strongly-coupled problem and accurate analytic results are not
possible. However, numerical data is available on the dispersion
spectrum in the N\'eel
phase\cite{trugman,ssneel,elser,poilblanc,reiter,mish}, and we know
that for a model with nearest-neighbor hopping, $t$, a single hole
has its dispersion minimum at the four $\vec{K}_p$ points shown in
Fig.~\ref{fig2}. We will proceed here using the results of a
self-consistent analysis of a single $f$ excitation above the U(1)
spin liquid state as carried out in previous work \cite{klr,suyu}.
This mean field analysis is reviewed in Appendix~\ref{ribhu} in our
notation. As we will see in Section~\ref{sec:neel}, the resulting
theory derived here predicts a hole dispersion in the N\'eel phase
consistent with the numerical
results\cite{trugman,ssneel,elser,poilblanc,reiter,mish}. We also
note that for suitable second-neighbor hopping, $t'$, the hole
dispersion in the N\'eel phase shifts its minima away from the
$\vec{K}_p$ points to $(\pi/a) (0, \pm 1)$, $(\pi/a)(\pm 1, 0)$:
this is believed to be relevant for the electron doped
cuprates\cite{tremblay}. The PSG analysis presented below will
require modifications of this case, but will not be discussed
here.

We already know that the fermionic holons, $f$, carry a U(1) gauge
charge $q= \pm 1$. The analysis in Appendix~\ref{ribhu} shows that
they also acquire an additional `valley' quantum number because
their dispersion minimum is not at the center of the Brillouin zone;
we will keep track of this at the cost of some additional
bookkeeping. There are 2 distinct valleys, with labels $v=1,2$, and
we choose a gauge in which their minima are at wavevectors
$\vec{K}_{1,2}$ shown in Fig.~\ref{fig2}. Summarizing, we have 4
species of spinless fermions $f_{qv}$, with U(1) charge $q=\pm 1$,
and valleys $v=1,2$. The PSG of the $f_{qv}$ can now be deduced from
the PSG listed in Table~\ref{table0}. We need to modify the
transformations here by additional factors arising from a factor
$e^{i \vec{K}_v \cdot \vec{r}}$, associated with shift of the
dispersion minimum to a finite wavevector. Such a procedure leads
the results in Table~\ref{tablepsg}.

Before we turn to the effective action for the $f_{qv}$, it is
useful to write down an expression for the physical electron
operator. This will allow to compute observable properties of our
theory, such as the location of the Fermi surfaces. This expression
can be obtained by requiring a spinon-holon composite to be gauge
invariant, and to transform under the PSG like a physical electron.
Because the $f_{qv}$ fermions reside near the $\vec{K}_p$ momenta
defined in Fig~\ref{fig2}, while the spinons are near zero momentum,
it is convenient to decompose the charge $-e$, spin-1/2 electron
annihilation operator $c_{\alpha} (\vec{r})$ into electron-like
components, $\Psi_{p\alpha}$, near $\vec{K}_p$:
\begin{equation}
c_{\alpha} (\vec{r})
 =
\sum_{p=1}^4 e^{i \vec{K}_p \cdot \vec{r}} \Psi_{p\alpha}
(\vec{r})\, .
\end{equation}
Now the PSG requirements yield unique bilinear combinations for the
$\Psi_{p \alpha}$
\begin{eqnarray}
\Psi_{1,3\alpha} &=& z_\alpha f_{+1}^\dagger \pm
\varepsilon_{\alpha\beta}
z^{\beta \ast} f_{-1}^\dagger \nonumber \\
\Psi_{2,4\alpha} &=& z_\alpha f_{+2}^\dagger \pm
\varepsilon_{\alpha\beta} z^{\beta \ast} f_{-2}^\dagger \, .
\label{psi}
\end{eqnarray}

We finally turn to the effective action for the $f_{qv}$ holons.
First, the terms bilinear in the $f$, invariant under the PSG, are:
\begin{equation}
\label{eq:Lf} \mathcal{L}_f =  \sum_{qv} f^\dagger_{qv}  \biggl(
\partial_\tau - i q A_\tau -\mu - \frac{(\partial_{\overline{j}} - i
q A_{\overline{j}} )^2}{2m_{v\overline{j}}} \biggr) f_{qv},
\end{equation}
where $\overline{j}$ extends over $\overline{x},\overline{y}$,
$m_{1\overline{x}}=m_{2\overline{y}}$ and
$m_{2\overline{x}}=m_{1\overline{y}}$ are the mass of the elliptical
hole pockets, $\mu$ is the hole chemical potential, and the
$\overline{x}$ and $\overline{y}$ directions are rotated by
45$^\circ$ from the principle square axes. The PSG prohibits a
linear derivative term in Eq.~(\ref{eq:Lf}), and so the $f_{qv}$
dispersion is an extremum at zero momentum; this pins the Fermi
surfaces to be centered at $\vec{K}_p$ in the N\'eel phase
(Fig.~\ref{fig2}), but not, as we will see, in the VBS phase.

Next, we include coupling between the holons and spinons invariant
under all PSG operations. There is an unimportant scalar coupling
$\sum_{\alpha q v} |z_\alpha|^2 f^\dagger_{qv} f_{qv}$, but the more
important term is
\begin{equation}
\mathcal{L}_c = i \widetilde{\lambda} \varepsilon^{\alpha\beta}
\left\{ f^\dagger_{+1} f_{-1} z_\alpha
\partial_{\overline{x}} z_\beta  + f^\dagger_{+2} f_{-2} z_\alpha
\partial_{\overline{y}}  z_\beta  \right\} + \mbox{c.c.},
\label{lc1}
\end{equation}
equivalent to the dipole coupling introduced by Shraiman and Siggia
\cite{ss}, arising from the hopping of electrons between
nearest-neighbor sites.

Finally, we have to consider couplings between the holons and the
U(1) gauge dynamics. The minimal coupling to the gauge field $A_\mu$
is already included in $\mathcal{L}_f$. Additionally, we have to
allow for couplings between the monopoles and the holons. These are
also specified by the PSG. First, we search for terms which involve
$\Phi_{\rm VBS}$ and a bilinear of the $f_{qv}$ fermions. However a
detailed analysis shows that there are {\em no such terms\/} which
are invariant under all the PSG operations; this is seen by first
listing the $f_{qv}$ bilinear invariants under $I_x^{\rm dual}$
(under which $\Phi_{\rm VBS}$ is invariant), and then noting that
their transformations under $R_{\pi/2}^{\rm dual}$ are incompatible
with those of $\Phi_{\rm VBS}$. Consequently, the coupling between
the VBS order and the fermions will be weaker than might have been
initially expected, and will vanish faster than $\langle \Phi_{\rm
VBS} \rangle \sim \Delta^{\beta_{\rm VBS}/\nu}$ as $\Delta
\rightarrow 0$. The simplest non-vanishing coupling turns out to
require the full $\Psi_p$ electron operators in Eq.~(\ref{psi}).
This has the form
\begin{eqnarray}
\mathcal{L}_{\rm VBS} &=& \lambda_{\rm VBS} \Phi_{\rm VBS}^\ast
\Bigl[ -i \bigl( \Psi^\dagger_1 \Psi_4 - \Psi^\dagger_4 \Psi_1  +
\Psi^\dagger_2 \Psi_3 - \Psi^\dagger_3 \Psi_2 \bigr) \nonumber \\
&~&~~~~~~~~~~~+ \bigl( \Psi^\dagger_1 \Psi_2 - \Psi^\dagger_2 \Psi_1
+ \Psi^\dagger_4 \Psi_3 - \Psi^\dagger_3 \Psi_4 \bigr) \Bigr] +
\mbox{c.c.} \label{lvbs}
\end{eqnarray}

Our complete field theory of the doped antiferromagnet is now
specified by the Lagrangian
\begin{equation}
\mathcal{L} = \mathcal{L}_J + \mathcal{L}_f + \mathcal{L}_c +
\mathcal{L}_{\rm VBS} \, .
\end{equation}
The following sections will describe its properties in the various
states in the limit of small hole density $\delta$.

\section{Doping the N\'eel state}
\label{sec:neel}

As noted earlier, there is much earlier work on the Fermi liquid
state obtained by doping a N\'eel
state\cite{trugman,ssneel,elser,klr,martinez,poilblanc,reiter,mish}.
Here we present an instructive argument which shows how these
results are reproduced in the present field-theoretic context. In
particular, we will show that the gauge charge index $q$ and the
spin quantum number are tied to each other in the N\'eel phase, and
also obtain the results quoted earlier on the volume of the hole
pockets.

Consider the $s<s_c$ N\'eel phase of $\mathcal{L}_z$ with $\langle
z_\alpha \rangle \neq 0$. By the Higgs mechanism, the condensate of
$z_\alpha$ renders the $A_\mu$ massive, and so we can safely
integrate the $A_\mu$ out. A crucial point is that the Higgs
mechanism also ties the U(1) gauge charge $q$ of the holons to the
spin quantum number $S_z$ along the N\'eel order (because of the
broken symmetry, $S_x$ and $S_y$ are not conserved); more precisely
$S_z = q/2$, and so the $f_{qv}$ carry all the quantum numbers of
the electron. By spin rotation invariance, we can always rotate the
$z_{\alpha}$ condensate (and without {\em any\/} rotation of the
spinless $f_{qv}$) to produce a N\'eel order $N^a = z^{\alpha \ast}
\sigma^{a\beta}_{\alpha} z_\beta$ (where $\sigma^a$ are the Pauli
matrices) polarized along the $(0,0,1)$ direction. Now examine the
response of the theory to a uniform magnetic field $H$ applied along
the $z$ direction. Under such a field, the only change in the action
is that \cite{qimp}
\begin{equation}
|(\partial_\tau
-iA_\tau)z_\alpha|^2 \mapsto ((\partial_\tau -iA_\tau -
H\sigma^z/2)z_\alpha ) \times ((\partial_\tau +iA_\tau +
H\sigma^z/2)z_\alpha^* )\, .
\end{equation}
Choosing $\langle
z_\alpha \rangle = \sqrt{|s-s_c|/u} \delta_{\alpha \uparrow}$, we
obtain a term in the Lagrangian
\begin{equation}
\mathcal{L} = \cdots (|s-s_c|/u) \bigl[ iA_\mu - (H/2)
\delta_{\mu\tau} \bigr]^2 + \cdots
\end{equation}
which gaps the $A_\mu$ photon. Note, however, that $A_\tau$
fluctuates about a non-zero value that is set by $H$. Integrating
out the $A_\mu$ and then evaluating $M_z = \left. \delta
\mathcal{S}/\delta H \right|_{H=0}$, we obtain an expression for the
magnetization density $M_z$
\begin{equation}
M_z = \frac{1}{2} \sum_{qv} q  f^\dagger_{qv} f_{qv} + \ldots
\end{equation}
where the ellipses represent an additional term which measures the
magnetization of the spin waves. This establishes, as claimed in
Fig.~\ref{fig1}, that $S_z = q/2$ for the fermions in the N\'eel
phase.

Now add a single hole and examine its coupling to the spin
excitations of the N\'eel state. The $A_\mu$ fluctuations (which
also represent spin waves in the N\'eel state) are expected to
strongly renormalize the effective mass of the hole. The
$\mathcal{L}_c$ term in Eq.~(\ref{lc1}) does not modify the fermion
dispersion at first order in $\widetilde{\lambda}$, but does lead to
a finite correction to the fermion mass at second order.
Consequently $f_{qv}$ fermions have a stable dispersion minimum at
zero momentum; or in other words, by Eq.~(\ref{psi}),  the
dispersion of the physical electrons has minima which remain pinned
at the $\vec{K}_p$. We will see that the influence of the
$\mathcal{L}_c$ term is quite distinct in the VBS state.

Now we consider the Fermi liquid state obtained with a total density
$\delta$ of the four fermion species $f_{qv}$. Each will form a
separate Fermi surface containing $\delta/4$ states. From
Eq.~(\ref{psi}) we deduce that there are 4 hole pockets centered at
the $\vec{K}_p$ wavevectors, each enclosing the area $\mathcal{A}_F
= (2\pi)^2 (\delta/4)$, as shown in Fig.~\ref{fig2}. The caption of
Fig.~\ref{fig2} shows that the same area is obtained in direct Fermi
liquid counting of electrons within the magnetic Brillouin zone.

\section{Doping the VBS state}
\label{sec:vbs}

This section will establish one of the new results of the paper: the
nature of the Fermi liquid state obtained by doping the VBS state.
We will show that the electronic quasiparticles have a dispersion
which is not centered in the $\vec{K}_p$, and consequently the
volume per hole pocket changes, as shown in Fig.~\ref{fig2}.

First, we recall some essential characteristics\cite{aa,rs} of the
U(1) spin liquid for $s>s_c$, $\delta=0$. This state has a $A_\mu$
``photon'', representing a single excitation associated with
rearrangements of valence bonds. The effective action of this photon
can be estimated in the large $N$ limit, where the spin index
$\alpha = 1 \ldots N$. In this limit we integrate out the $z_\alpha$
quanta in the action $\mathcal{L}_J$, and obtain an effective for
the $A_\mu$. Details appear in Appendix~\ref{cpn}, and at scales
larger than the spin correlation length,  $\sim 1/\Delta$, we have
the Lagrangian
\begin{equation}
\mathcal{L}_A = \frac{N}{48 \pi \Delta} \left(
\epsilon_{\mu\nu\lambda} \partial_\nu A_\lambda \right)^2 \label{la}
\end{equation}
As shown in Appendix~\ref{nrt}, in the non-relativistic limit of
slowly moving excitations, this photon mediates a logarithmic
``Coulomb'' interaction $V(r)$ between charged
particles\cite{wensc,gm}:
\begin{equation}
V(r) = \frac{12 \Delta}{N} \ln \left( r \Delta \right) \, ,
\label{vra}
\end{equation}
which also appears in Eq.~(\ref{vr}). The divergence of $V(r)$ as $r
\rightarrow \infty$ implies that only configurations which are net
charge zero have finite energy. Of course, at sufficiently large
$r$, we also have to consider the modifications due to
monopoles\cite{rs,senthil1}, which change the $A_\mu$ mediated
interaction from logarithmic to a linearly confining one. However,
this does not happen until a confinement length scale denoted
$\xi_{\rm VBS}$ in Ref.~\onlinecite{senthil1}, whose divergence as
$\Delta \rightarrow 0$ is determined by the scaling dimension of the
four-monopole operator, so that $\xi_{\rm VBS} \gg 1/\Delta$. So
there is substantial scale over which monopole effects can be
neglected, and we can work with the $V(r)$ in Eq.~(\ref{vr}).

Now let us examine the spectrum of a single $f$ hole in the VBS
state. We will see that its key quantum numbers are determined on a
scale smaller than $\xi_{\rm VBS}$, and so we can work with the
non-compact gauge field obeying the action in Eq.~(\ref{la}). As
shown in Appendix~\ref{nrt}, the coupling to $A_\tau$ causes this
holon to have an `electrostatic' self energy which diverges
logarithmically with system size \cite{qimp}, and so a spin-singlet
single holon state is not stable. Rather, the holon will peel off a
single spinon from above spinon gap, and form a $S=1/2$, charge $e$
bound state \cite{qimp}. This bound state is electron-like and is
neutral under the $A_\mu$ U(1) gauge force. A finite density of such
bound states can then form a Fermi surface with charge $e$, $S=1/2$
quasiparticles. We also have to consider the pairing between holons
with opposite U(1) gauge charges, induced by the $A_\mu$ gauge force
(this attractive force must be balanced against the repulsive
Coulomb force associated with the physical electromagnetic charge
$e$ of each holon). This is expected to lead to superconductivity at
low enough temperatures, but we will not discuss this here; our
focus is on the normal state.

We now discuss the wavefunction and quantum numbers of the
holon-spinon bound state. This can be addressed in a
non-relativistic theory of holons and spinons interacting via a
Coulomb force, as discussed in Appendix~\ref{nrt}. The force in
Eq.~(\ref{vr}) will bind the spinons/antispinons with the holons
into gauge neutral combinations. For each holon species $f_{qv}$,
there are {\em two\/} such bound states, because the spinon can have
spin up or down. So there are a total of 8 electron-like bound
states: this is the crucial difference from the electron-like bound
states in the N\'eel state, where there were only 4 electron-like
quasiparticles. Another key difference is that the Shraiman-Siggia
term, $\mathcal{L}_c$, mixes these 8 bound states, with significant
consequences for the structure of the Fermi surfaces.

The form of $\mathcal{L}_c$ in Eq.~(\ref{lc1}) makes it clear that
the bound state of a $f_{+1}$ holon with a spinon will be mixed with
the bound state of a $f_{-1}$ holon with a spinon (parallel
considerations apply to the $f_{\pm 2}$ holons). Before writing the
Schr\"odinger equation for this bound state, we need to decompose
the relativistic field $z_\alpha$ into non-relativistic, canonical
boson fields $p_\alpha$ (the spinon) and $h^\alpha$ (the
anti-spinon). At low momenta, this is done by the parameterization
(as in Eq.~(\ref{nr1}))
\begin{equation}
z_\alpha = \frac{1}{\sqrt{2\Delta}} \left( p_\alpha +
\varepsilon_{\alpha\beta} h^{\beta\dagger} \right) \label{znr0}
\end{equation}

We now consider the two-component wavefunction describing the
$\mathcal{L}_c$ induced mixing between the bound state of $f_{+1}$
and $h_\alpha$ and the bound state of $f_{-1}$ and $p_\alpha$. Let
us label the co-ordinates of the holon by $\vec{r}_h$ and that of
the spinon by $\vec{r}_s$. Then the bound state wavefunction has a
doublet structure between these states
\begin{equation}
\Phi (\vec{r}_s, \vec{r}_h) = e^{i \vec{P} \cdot \vec{R}} \left( \begin{array}{c} \phi_+ (\vec{r}) \\
\phi_- (\vec{r})
\end{array} \right)
\label{Phiphi}
\end{equation}
and is expressed in terms of the center-of-mass and relative
co-ordinates
\begin{eqnarray}
R_{\overline{j}} &=& \frac{m_{v\overline{j}} r_{h\overline{j}} +
\Delta r_{s\overline{j}}}{m_{v\overline{j}} + \Delta}
\nonumber \\
\vec{r} &=& \vec{r}_h - \vec{r}_s,
\end{eqnarray}
where $\Delta$ is the mass of the spinon in the non-relativistic
limit, and $m_{vj}$ are the holon masses in $\mathcal{L}_f$.
Similarly, we can also define the total masses and the reduced
masses
\begin{eqnarray}
M_{\overline{j}} &=& m_{v\overline{j}} + \Delta \nonumber \\
\rho_{\overline{j}} &=& \frac{m_{v\overline{j}}
\Delta}{m_{v\overline{j}} + \Delta}
\end{eqnarray}

For completeness, we also note the operator representation of this
state:
\begin{eqnarray}
|\alpha \vec{P} + \rangle &=& \int d\vec{r_S}d\vec{r_h} e^{i \vec{P}
\cdot \vec{R} }
\phi_+ (\vec{r}) f^\dagger_{+1}(\vec{r}_h) h^{\alpha\dagger} (\vec{r}_s) | 0 \rangle\nonumber \\
|\alpha \vec{P} - \rangle &=& \int d\vec{r_S}d\vec{r_h} e^{i \vec{P}
\cdot \vec{R} } \phi_- (\vec{r}) f^\dagger_{-1} (\vec{r}_h)
p^{\alpha\dagger}(\vec{r}_s) | 0 \rangle \label{ket}
\end{eqnarray}

Using a reduction closely related to that presented in
Appendix~\ref{nrt}, we deduce that the Schr\"odinger equation obeyed
by $\phi_\pm (\vec{r})$ is then
\begin{eqnarray}
&& \left( \frac{P_{\overline{j}}^2}{2 M_{\overline{j}}} -
\frac{\partial_{\overline{j}}^2}{2\rho_{\overline{j}}} + V(r)
\right)\left( \begin{array}{c} \phi_+ (\vec{r}) \\
\phi_- (\vec{r})
\end{array} \right) \nonumber \\
&&~~ - \left(\frac{\widetilde{\lambda}}{2 \Delta} \left[
\frac{2\Delta P_{\overline{x}} }{M_{\overline{x}}} \delta^2 (r) + i
\delta^2 (r)
\partial_{\overline{x}} + i \partial_{\overline{x}} \delta^2 (r)\right]
\right) \left( \begin{array}{c} \phi_- (\vec{r}) \\
\phi_+ (\vec{r})
\end{array} \right) = E_{\vec{P}} \left( \begin{array}{c} \phi_+ (\vec{r}) \\
\phi_- (\vec{r})
\end{array} \right) \label{se}
\end{eqnarray}
We first examine this equation at $\lambda=0$. For small spin gap
$\Delta$, the U(1) gauge potential potential in Eq.~(\ref{vr}),
$V(r) = (12 \Delta/N) \ln (r \Delta)$, binds the holons and spinons
over a length scale $\sim 1/\Delta$, with $\phi_{\pm} (0) = \phi (0)
\sim \Delta$. It is also clear that $\vec{P}$ dependence of
$E_{\vec{P}}$ is only in the center-of-mass kinetic energy. The
specific information we need from Eq.~(\ref{se}) is the modification
of the dispersion $E_{\vec{P}}$ induced by $\mathcal{L}_c$. This can
be estimated in perturbation theory in $\widetilde{\lambda}$ using
the matrix element $\langle \alpha \vec{P} + | \mathcal{L}_c|\alpha
\vec{P} -\rangle = -\widetilde{\lambda}|\phi(0)|^2
P_{\overline{x}}/M_x$ (we used the fact that $\partial_x\phi(0)=0$),
and leads to
\begin{equation}
E_{\vec{P}} = E_0 + \frac{P_{\overline{j}}^2}{2 M_{\overline{j}}}
\pm \frac{\widetilde{\lambda} P_{\overline{x}} |\phi (0)|^2}{
M_{\overline{x}}} \label{eqshift}
\end{equation}
with eigenmodes which correspond precisely to the electron states in
Eq.~(\ref{psi}). So the bound state dispersion has a minimum at
$K_{\overline{x}} = \pm \lambda$ where $\lambda =
\widetilde{\lambda} |\phi (0)|^2 \sim \Delta^2$. This is responsible
for the shift in the centers of the elliptical hole pockets shown in
Fig 2.

The usual counting argument now allows us to deduce the area of each
hole pocket. There are 4 inequivalent pockets, and a factor of 2
degeneracy for spin; so $\mathcal{A}_F = (2 \pi)^2 (\delta /8)$. The
factor of 2 difference from the result obtained in the N\'eel phase
is one of our key results.

The state we have described so far is actually {\em not\/} a
conventional Fermi liquid: the area enclosed by its Fermi surface is
not the same as that of a non-interacting electron gas at the same
filling and with the same size of unit cell. In the non-interacting
case one would have a Fermi surface area $(2\pi)^2(1-\delta)/2$. In
our case, where the holes were doped on a background spin liquid
state with a gapless photon excitation, we obtain that the
electron-like area enclosed by the Fermi surface is $(2\pi)^2
(1-\delta/2)$ . This state is a {\em fractionalized Fermi liquid\/}
\cite{ffl}, obtained here in a single band model, in contrast to its
previous appearance in Kondo lattice models.

The instability of the spin liquid to confinement and VBS order
induced by monopoles \cite{rs,senthil1}, and the associated halving
of the Brillouin zone will finally transform the state into one
obeying the conventional Luttinger theorem, with an electron-like
area enclosed by the Fermi surface of $(2\pi)^2 (1 - \delta)/2$.
There will be Bragg reflection across the reduced Brillouin zone
boundaries.  However, because of the shift in the minimum of the
holon-spinon bound state dispersion discussed above, this mixing is
negligible as long as $\sqrt{\delta} < \lambda \sim \Delta^2$, and
so can always be neglected as $\delta \rightarrow 0$. This
establishes the Fermi surface structure claimed in Fig.~\ref{fig2}.

At larger $\delta$, the hole pockets will cross the dashed lines in
Fig.~\ref{fig2}, and Bragg reflections will split the Fermi
surfaces. We can analyze this magnitude of the mixing matrix element
in the Hamiltonian,  using $\mathcal{L}_{\rm VBS}$ in
Eq.~(\ref{lvbs}). Using Eqs.~(\ref{znr0}) and (\ref{Phiphi}), we
obtain a matrix element of order
\begin{equation}
\lambda_{\rm VBS}  \langle \Phi_{\rm VBS} \rangle \frac{|\phi
(0)|^2}{\Delta} \sim \lambda_{\rm VBS} \Delta^{1 + \beta_{\rm
VBS}/\nu}.
\end{equation}
As expected, this does vanish faster than $\Delta^{\beta_{\rm
VBS}/\nu}$, and is responsible for the weak Bragg reflection across
the reduced Brillouin zone boundaries of the VBS state in
Fig.~\ref{fig2}. The splitting of the Fermi surface is negligible
provided the matrix element is smaller than the hole kinetic energy,
or  $ \delta > \Delta^{1 + \beta_{\rm VBS}/\nu}$. Over such a
possible regime of larger $\delta$, the basic pictures of
Figs.~\ref{fig1} and~\ref{fig2} continue to hold. The photoemission
intensity of the ``shadow'' Fermi surfaces noted in Fig.~\ref{fig2}
is proportional to the square of the matrix element or $\sim
\Delta^{2(1 + \beta_{\rm VBS}/\nu)}$.

\section{Holon metal and quantum criticality}
\label{sec:qc}

To complete the picture, let us address the physics of the shaded
region in Fig.~\ref{fig1}. Neglecting the gauge fluctuations, the
holons and spinons are independent. The four holon species then form
their own Fermi-surfaces centered around $K_{1,2,3,4}$. In the
N\'eel phase, the spinons condense leading to long-range order. In
this case the physical electron fields {\em are} the holons  On
increased doping, N\'eel order is destroyed, which corresponds to
the development of a gap for the spinons. Now we have a finite
density of holons that we expect will form a `holon metal' with a
holon Fermi surface. Including gauge fluctuations we have a theory
of a finite density of holons $f_{qv}$ interacting with spinons
$z_\alpha$ via a U(1) gauge force $A_\mu$. At sufficiently long
scales, the holons will `Thomas-Fermi' screen that longitudinal
$A_\tau$ force\cite{hermele}, and so obviate the binding into gauge
neutral combinations. Consequently the spinons and holons remain as
relatively well-defined excitations, and we expect to enter a
fractionalized holon metal phase. This screening can be prevented if
the spacing between the holons ($\sim 1/\sqrt{\delta}$) is larger
than the holon-spin binding length $\sim 1/\Delta$, or $\delta <
\Delta^2$; this fixes the boundary of the shaded region in
Fig.~\ref{fig1}. Alternatively, the same results follows from our
argument in Section~\ref{sec:intro} that the scaling dimension of
$\delta$ is 2. The shaded region will exclude the unshaded region of
the VBS phase where the Fermi surface splitting is negligible
(discussed in the previous paragraph) provided $\beta_{\rm VBS}>
\nu$, an inequality that holds at least for large $N$. Note that it
is the competition between the holon kinetic energy and the gauge
mediated attractive interaction with the spinons that determines the
relative stability of the holon metal and the VBS phases.

It is this fractionalized phase that we propose as a candidate for
the pseudo-gap phase of the cuprates. Due to the presence of the
gapless gauge field the low energy properties of the holon metal
will be modified in well-known ways from that of a non-interacting
holon picture, and we will not repeat these further here. We
emphasize again here that the ``holon-metal'' phase should be stable
to confinement due to monopoles following previous screening
arguments~\cite{hermele}. The metallic holon-metal phase will
however inevitably be unstable at low-$T$ to pairing and hence
superconductivity.

We now describe the criticality of the hole spectrum at $s=s_c$. We
assume here an observation scale (frequency ($\omega$) or
temperature ($T$)) large enough, or a $\delta$ is small enough, so
that the holes can be considered one at a time. A key observation
about a single hole is that its quadratic dispersion ({\em i.e.\/}
the terms proportional to $1/(2 m_{v\overline{x},\overline{y}})$ in
$\mathcal{L}_f$) is an irrelevant perturbation on the quantum
critical point of $\mathcal{L}_z$ which involves excitations which
disperse linearly with momentum; a similar observation was made in
Ref.~\onlinecite{ssvojta} for Landau-Ginzburg-Wilson critical point
of O(3) model. Consequently, the hole may be considered localized,
and its physics is closely related to the single impurity in a spin
liquid problem analyzed in Ref.~\onlinecite{qimp}. Here we are
interested in the single hole Green's function and this requires the
overlap between states of the spin liquid with and without the
impurity. This can be computed by analyzing the quantum critical
theory of $\mathcal{L}_z$ coupled to one holon localized at $r=0$,
represented by the $r$-independent Grassmanian $f(\tau)$:
\begin{displaymath}
\mathcal{S}_{qc} = \int d^2 r d\tau \mathcal{L}_z + \int d \tau
f^\dagger \left( \frac{\partial}{\partial \tau} + \varepsilon_0 - i
A_\tau (r=0,\tau) \right) f.
\end{displaymath}
Here $\varepsilon_0$ is an arbitrary energy fixing the bottom of the
holon band, and, following earlier arguments \cite{qimp}, the only
relevant coupling between the `impurity' holon degree of freedom and
the bulk degrees of freedom of $\mathcal{L}_z$ is the gauge coupling
associated with $A_\tau$. The spectral function of a physical charge
$e$, $S=1/2$ hole is given by the two-point correlation of the
composite operator $z_\alpha f^\dagger$. If this operator has
scaling dimension $\eta_h/2$, then the universal critical hole
Green's function, $G_h$, is independent of wavevector and obeys the
scaling form
\begin{equation}
G_h = T^{-(1-\eta_h)} \Phi (\hbar (\omega-\varepsilon_0)/k_B T)
\label{gh}
\end{equation}
where $\Phi$ is a universal scaling function. At $T=0$, we obtain an
incoherent spectrum associated with the power-law singularity $G_h
\sim (\omega - \epsilon_0)^{-1+\eta_h}$. We have computed $\eta_h$
by a standard $1/N$ expansion for $G_h$ under the action
$\mathcal{S}_{qc}$; the computation is described in
Appendix~\ref{qimp}, and the result is
\begin{equation}
\eta_h = 1 - \frac{36}{N\pi^2} + \mathcal{O} \left( \frac{1}{N^2}
\right) \label{etah}
\end{equation}

Another perspective on the exponent $\eta_h$ is obtained by mapping
it to an observable in the lattice non-compact $\mathbb{CP}^1$ model
studied by Motrunich and Vishwanath \cite{mv}. On a
three-dimensional cubic lattice with spacetime points $j$, the model
has the complex spinor fields $z_{j\alpha}$ on each lattice site,
and a vector potential $A_{j\mu}$ on each link extending along the
$\mu$ direction from site $j$. The hole Green's function at
imaginary time $M_\tau$ (in units of the lattice spacing) is then
given by the two-point correlator of the $z_{j\alpha}$ in the time
direction along with an intermediate `Wilson line' operator
(representing the contribution of the $f$ fermion) which renders the
correlator gauge invariant (see also Refs.~\onlinecite{wen,jinwu}):
\begin{displaymath}
G_h (M_\tau) = \left\langle z^{\alpha \ast}_{j} \exp \left( i
\sum_{n=0}^{M_\tau-1} A_{j+n\hat{\tau},\tau} \right) z_{j+M_\tau
\hat{\tau},\alpha} \right\rangle_{\rm NCCP};
\end{displaymath}
this is expected to decay as $e^{- \widetilde{\varepsilon}_0 M_\tau}
/M_\tau^{\eta_h}$ for large $M_\tau$ at the quantum critical point.
The continuum limit of the above correlator was computed by Kleinert
and Schakel \cite{kleinert} in the $1/N$ expansion: their result
agrees with Eq.~(\ref{etah}).

\subsection{Photoemission spectra}
\label{sec:photo}

The result Eq.~(\ref{gh}) describes universal component of the
electronic spectrum in the limit where ``irrelevant'' terms such as
fermion dispersion and $\mathcal{L}_c$ in Eq.~(\ref{lc1}) are
neglected. With an eye towards comparison with experiments
\cite{mohit}, this subsection will present numerical results in the
full Brillouin zone for a range of $T$ with these effects taken into
account.

Formally similar computations have been carried out by Weng {\em et
al.}\cite{weng} at $T=0$. However, they applied their result to the
N\'eel phase, whereas we have argued that the N\'eel phase has a
conventional Fermi liquid character. They also do not obtain the
shift of the spectral weight maxima away from the
$\vec{K}_{1,2,3,4}$ points in the Brillouin zone.

We are interested in the electron spectral function that would be
measured in a photo-emission experiment. The measured photo-emission
intensity is related to the spectral density of the physical
electron Greens' function.
\begin{equation}
\label{cgrdef}
  G_{\alpha\beta}^c(r,\tau)= -\langle \mathcal{T}[c_{r\alpha}(\tau) c_{\beta}^\dagger(0) ]\rangle
\end{equation}
In order to re-write this expression in terms of the $z$ and $f$
fields we need an expression for the physical electrons in terms of
these fields.  As shown earlier, we can derive such a relation from
the projective symmetry group (PSG) transformation properties of the
various fields Eq.~(\ref{psi}),
\begin{eqnarray}
c_{r\alpha} &=& e^{iK_1\cdot r}[z_\alpha f^\dagger_{+1}+
\varepsilon_{\alpha\beta} z^{\beta\ast} f^\dagger_{-1}]\nonumber\\
&+& e^{iK_3\cdot r}[z_\alpha f^\dagger_{+1}- \varepsilon_{\alpha\beta} z^{\beta\ast} f^\dagger_{-1}]\nonumber\\
&+& e^{iK_2\cdot r}[z_\alpha f^\dagger_{+2}+ \varepsilon_{\alpha\beta} z^{\beta\ast} f^\dagger_{-2}]\nonumber\\
&+& e^{iK_4\cdot r}[z_\alpha f^\dagger_{+2}-
\varepsilon_{\alpha\beta} z^{\beta\ast} f^\dagger_{-2}]
\end{eqnarray}
Using this expression, we can calculate the contributions to
Eq.~(\ref{cgrdef}) under the
$\mathcal{L}=\mathcal{L}_z+\mathcal{L}_f+\mathcal{L}_c$.

In a ``cartoon'' mean-field state for the holon-metal phase we
neglect the gauge fluctuations. The holons and spinons are
independent (we temporarily also set $\widetilde{\lambda}=0$, but
shall calculate the effect of its presence shortly). As described
above in the Neel state the physical electron fields {\em are} the
holons and thus in this phase the the photo-emission response should
have coherent peaks exactly where the holons do, i.e. $(\pm\pi/2,\pm
\pi/2)$. This is consistent with photo-emission experiments in the
doped cuprates with long-range N\'eel order present\cite{kim}. On
increased doping and entering the `holon-metal phase with a finite
spin gap, the spinons are not condensed and the associated spinonic
fluctuations play a crucial role in the measured electronic spectra.
We will calculate the spectra within this mean field picture and
simply convolve the free spinon and holon Greens functions to get
the electron Greens function. Qualitatively the neglected gauge
fluctuations are expected to have two effects. First the scattering
of the gapless holons with the gapless transverse gauge fluctuations
leads to a singular frequency dependent self-energy for the holon
Greens function. Upon convolving with the spinon this tends to make
the hole spectrum more incoherent than the mean field result. Second
as discussed extensively in previous sections the gauge field leads
to attraction between the holons and spinons. This gauge-induced
attractive interaction tends to make the hole spectrum more coherent
than the mean field result. These two kinds of effects may be viewed
as involving self-energy and vertex corrections due to gauge
interactions to the diagrams that contribute to the electron Greens
functions. It is expected that these effects do not qualitatively
change the mean field results to be presented below, at least at the
high temperatures of interest.

\begin{figure}
  \includegraphics[width=3in]{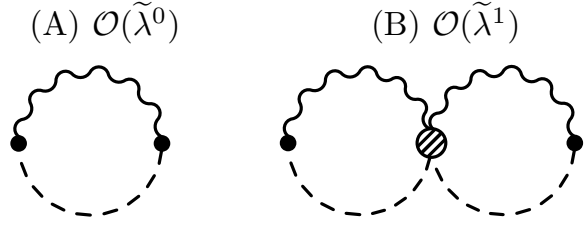}
  \caption{Diagrams contributing to $A^c(k,\omega)$ at (A) zeroth order
  and (B) first order in Eq.~(\ref{lc1}). The wavy (dashed) lines are
  the spinon (holon) propagators. The shaded vertex corresponds to $\widetilde{\lambda}$.}
  \label{fig:diag}
\end{figure}
Turning to the calculation of the spectral function in the cartoon
holon-metal phase described above, we find good comparison with
recent temperature dependent photo-emission experiments in the
normal ``pseudo-gap'' state of the cuprates~\cite{mohit}. In the
cartoon state the propagators for the $z$ and $f$ field are,
$G^z_{k,i\nu_m}\equiv-\langle z z^\dagger \rangle =-\frac{1}{\nu_m^2
+ E^2_k}$ and $ G^f_{k,i\omega_n}\equiv- \langle f f^\dagger
\rangle= \frac{1}{i\omega_n - \epsilon_k}$, where $E_k$ is the
spinon dispersion and $\epsilon_k$ is the kinetic energy of the
holons.

Neglecting fluctuations of the gauge field and setting
$\widetilde{\lambda}=0$, Eq.~(\ref{cgrdef}) becomes after
Fourier-transformation,
\begin{eqnarray}
\label{eq:speczro} G^c_{k,i\omega_n}= 2 \sum_{P}\int
\frac{d^2q}{(2\pi)^2} T
\sum_{m}G^z_{q,i\nu_{m}}G^f_{q-k+K_P,i\nu_{m}-i\omega_n}.
\end{eqnarray}
The spectral function of the electron is thus a convolution of the
spectral functions of its composite ``holon'' and ``spinon''
particles, diagrammatically this may be simply represented as
Fig.~\ref{fig:diag}(A).

We can now easily complete the Matsubara sums and analytically
continue $i\omega_n \rightarrow \omega+i\eta$, to obtain the
expression for the spectral density $A=-2 \Im [G^R]$. Note that the
spectral functions do not satisfy the usual sum rule $\int
\frac{d\omega}{2\pi}A(\omega)=1$ for fermions. This is because we
are working the in projective subspace of the $t$-$J$ model.

\begin{figure}[!t]
\includegraphics[width=7in]{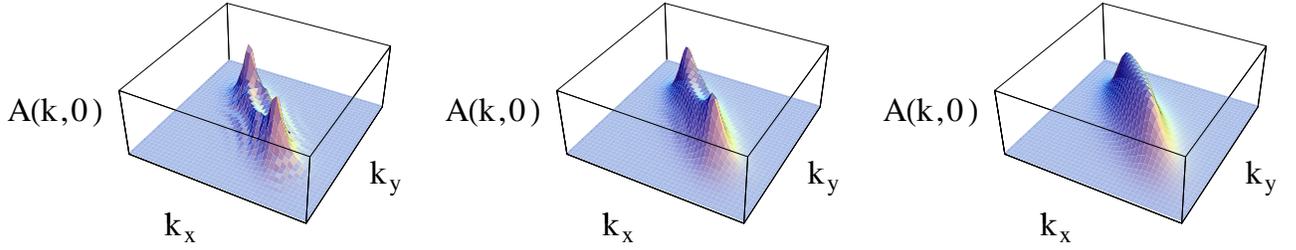}
\caption{The change in the spectral response as the parameter $\eta$
(spectral broadening) is increased. In units of the spinon gap
$T=0.5$. From left to right $\eta=0.2,0.4,0.8$. In the limit of
perfect resolution the response consists of ellipses centered around
the $\vec{K}_p$. The spectral peaks are due to the higher density of
states perpendicular to the zone diagonals due to the larger holon
mass in this direction. The rest of the results shown in this paper
have $\eta=0.8$ to model the experimental resolution. }
\vspace*{-8pt} \label{spec_res}
\end{figure}

\begin{figure}[!t]
\includegraphics[width=6in]{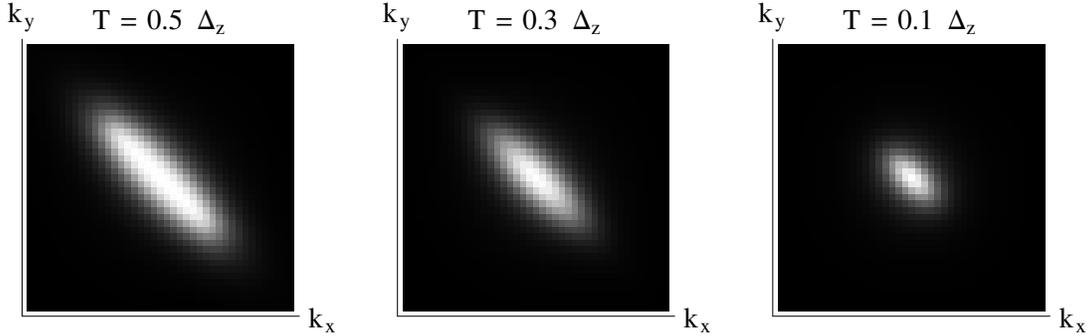}
\caption{Evolution of $A^c_{\rm HM}(k,0)$ as a function of $T$ in
the first quadrant of the square lattice BZ, calculated with
Fig.~\ref{fig:diag}A. These results bear close resemblance to the
shrinking ``Fermi Arcs'' seen in photo-emission experiments on the
underdoped cuprates~\cite{mohit}.  The numerical evaluation of
Eq.~(\ref{eq:cspec_arcs}) was made with
$m_{1\overline{x}}/m_{1\overline{y}}=0.05$ to model the anisotropic
holon dispersion. These results are in the regime where the spectral
resolution $\eta>T$, as in current experiments~\cite{mohit}. Note
that this figure has $\widetilde{\lambda}=0$ and the ``Fermi arcs''
shrinks to $(\frac{\pi}{2},\frac{\pi}{2})$ at $T=0$.} \vspace*{-8pt}
\label{cspec_arcs}
\end{figure}

Let us consider $\omega=0$ in the ``holon-metal'' phase, then
Eq.~(\ref{eq:speczro}) becomes,
\begin{equation}
\label{eq:cspec_arcs} A_{\rm HM}^c(k,0)= 2\pi \frac{1}{V}\sum_{p,v}
\frac{\delta(\epsilon-E)}{E}[n_B(E)+n_F(\epsilon)],
\end{equation}
where we have used $\epsilon=\epsilon_{p-k+K_P}$ and $E=E_p$ for
shorthand. Note that at $T=0$ this evaluates to zero and thus these
contributions correspond to propagation of physical electrons in the
presence of thermally excited spinons and holons. Using $\epsilon_k
=
\frac{k^2_{\overline{x}}}{2m_{v\overline{x}}}+\frac{k_{\overline{y}}^2}{2m_{v\overline{y}}}$
(for simplicity we set the chemical potential of the holons $\mu=0$)
and $E_k=\sqrt{\Delta_z^2 +c^2k^2}$, we can compute the zero energy
spectral weight in $k$-space. We expect that $A(k,0)$ should have
weight only for those momenta, $k$ at which a relative shift by $k$
of the spinon dispersion with respect to the holon dispersion causes
the bottom of the gapped spinon dispersion to have the same energy
as the holon. We thus expect spectral weights on ellipses centered
around the $K_p$ points. We note however that current experiments
are carried out in the regime where the spectral resolution
(16-20meV) is larger than $T$ (110-200 K)~\cite{mohit}. For this
reason, the ellipses predicted by our theory can not be resolved, as
illustrated in Fig.~\ref{spec_res}. To make explicit contact with
the experiments, we shall henceforth work in the regime where
$\eta>T$ ($\eta$ is the imaginary part used in the analytic
continuation).

The results are displayed in Fig.~\ref{cspec_arcs} as a function of
the temperature $T$, they have many features similar to the spectral
function measured in recent photo-emission experiments~\cite{mohit}.
At finite-$T$, there are four ``Fermi arcs'' which have lengths that
shrink with lowering $T$. In the holon-metal theory this shrinkage
is due to that fact that as $T$ is lowered there are fewer and fewer
thermally excited spinons present that are required to have an
electronic excitation. An important difference between the
experiments and the results presented in Fig.~\ref{cspec_arcs} is
that in the photo-emission experiments in the pseudo-gap phase, the
point to which the ``Fermi arcs'' shrink {\em is shifted away} from
the $\vec{K}_p$. We now show that this observation follows on
including the effect of Eq.~(\ref{lc1}).

We have so far excluded the ``Shraiman-Siggia'' term that couples
the holons and the spinons, Eq.~(\ref{lc1}), i.e., we have used
$\widetilde{\lambda}=0$. The first order correction to the $A^c$ due
to Eq.~(\ref{lc1}) is given by the diagram shown in
Fig.~\ref{fig:diag}(B).
\begin{figure}[!t]
  \includegraphics[width=6in]{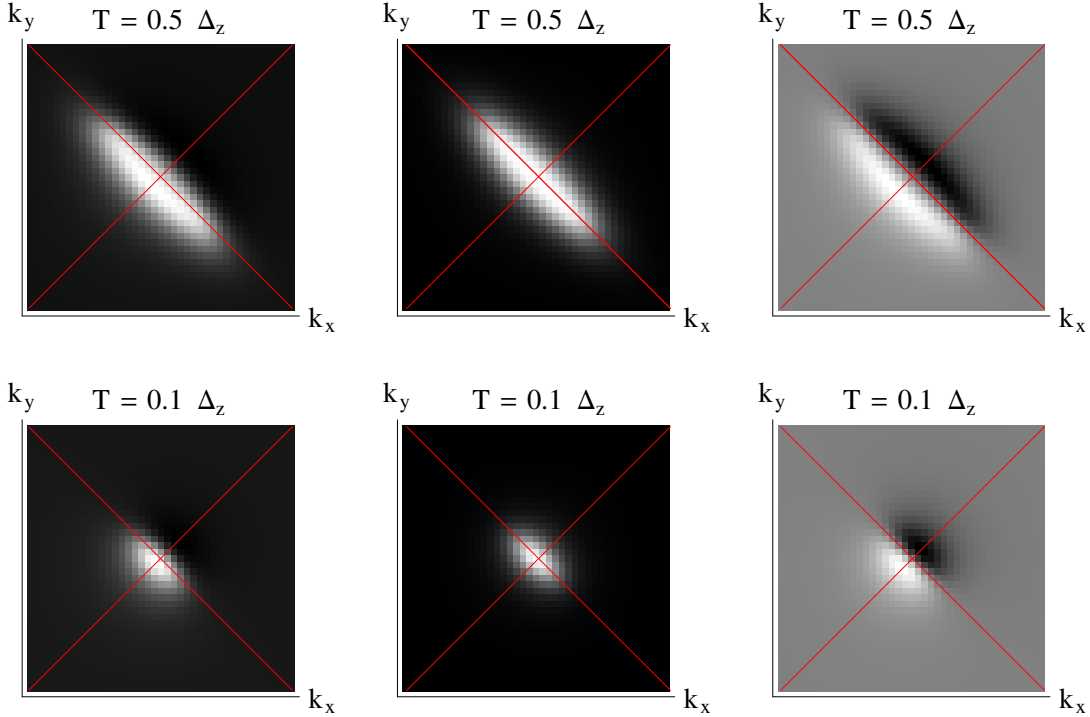}
\caption{(color online) The first column shows the shift of the
``nodal point'' of Fermi arcs from the $\vec{K}_p$ on inclusion of
the effect of Eq.~(\ref{lc1}) at first order in perturbation theory,
Fig.~\ref{fig:diag}B; the plot is of $A^c=A^{c(0)}+A^{c(1)}$. The
next two show the individual contributions, $A^{c(0)}$ and
$A^{c(1)}$. The second row is with the same physical parameters as
the first set but at a lower $T=0.1 \Delta_z$. All these data are in
the regime where $\eta>T$. Lines joining $(0,0)$ to $(\pi,\pi)$ and
$(0,\pi)$ to $(\pi,0)$ have been included for clarity.}
\vspace*{-8pt} \label{fig:folambda}
\end{figure}
We now focus on the momenta close to
$K_1=(\frac{\pi}{2},\frac{\pi}{2})$ and hence drop the three other
square lattice symmetry related terms. After Fourier transformation
we obtain,
\begin{eqnarray}
\label{eq:spec1}
G^{c{(1)}}_{k+K_1,i\omega_n}&=&\frac{2i\widetilde{\lambda}}{V^2}\left[2\sum_{p,q}i(k+p)_xf(p,q,k,i\omega_n)
\right]
\end{eqnarray}
where,
\begin{eqnarray}
f(p,q,k,i\omega_n)&=& g(p,k,i\omega_n)g(q,k,i\omega_n),\nonumber\\
g(p,k,i\omega_n)&=&T\sum_{n_1}G^f_{p,i\omega_{n_1}}G^z_{k+p,i\omega_n+i\omega_{n_1}}
\end{eqnarray}
In order to calculate the spectral function we need to complete the
Matsubara sums and then continue $i\omega_n \rightarrow \omega + i
\eta$ and get the imaginary part of $f$. Note that because one can
always put $p\rightarrow -p$ in the sums above the spectral function
contribution changes sign when $k_x\rightarrow -k_x$, i.e. depending
on the sign of $\widetilde{\lambda}$, it shifts the spectral weight
in $A^c(k,0)$ towards(away) from the $\Gamma$-point along the zone
diagonal.

For $\omega=0$, we obtain for the function $g(p,k,\omega=0)$,
\begin{eqnarray}
\Re[g]&=&\frac{1}{2E}\left[\frac{n_B(E)-n_F(\epsilon)+1}{E+\epsilon}-\frac{n_B(E)+n_F(\epsilon)}{E-\epsilon}\right]\nonumber \\
\Im[g]&=&-\frac{\pi}{2E}[n_B(E)+n_F(\epsilon)]~~\delta(E-\epsilon).
\end{eqnarray}
where we have uses the short-hand, $E=E_{k+p}$ and
$\epsilon=\epsilon_p$.

We now evaluate the sums in Eq.~(\ref{eq:spec1}) numerically.
Results are presented in Fig.~\ref{fig:folambda}. Inclusion of the
$\widetilde{\lambda}$ term does not change the general features of
the shrinking ``Fermi arcs'' observed in Fig.~\ref{cspec_arcs},
instead it shifts the nodal point away from the $\vec{K}_p$, just as
is seen in the experiments. It should be noted that the holons still
have their dispersion minima at the $\vec{K}_p$, only the physical
electrons have their spectral weight shifted away from $\vec{K}_p$
because of $\widetilde{\lambda}$. This should be contrasted with the
long-range ordered N\'eel phase where even Eq.~(\ref{lc1}) cannot
change the pinning of the center of the spectral weight away from
the $\vec{K}_p$ (as observed in experiments\cite{kim}).

\section{Conclusions}
\label{sec:conc}

The starting point of this paper was the deconfined quantum critical
point between the N\'eel and VBS ground states of a $S=1/2$ square
lattice antiferromagnet. We added a small density of holes to both
states and found distinct Fermi liquid states. Their difference was
characterized in Eq.~(\ref{ratio}), and ultimately had its origin in
the following results: ({\em i\/}) the dispersion of a single hole
in a N\'eel state can have the minimum of its dispersion pinned at
the symmetrical points $\vec{K}_p$ (Fig~\ref{fig2}) over a finite
range of parameters; ({\em ii\/}) the minimum of the dispersion of a
single hole in the VBS state is generically shifted away from the
$\vec{K}_p$, as established in Eq.~(\ref{eqshift}); the binding of
the holon and spinon into a charge $e$ $S=1/2$ quasiparticle in the
VBS phase was crucial in obtaining this shift. We also argued that
at finite hole doping, $\delta$, these Fermi liquid states were
likely not connected by a direct quantum phase transition, but that
the insulating deconfined quantum critical point opened up into a
non-Fermi-liquid holon metal phase.

We also discussed the finite temperature, ``quantum critical'',
holon metal phase obtained above the low-temperature ordered phases.
The spectra shown in Section~\ref{sec:photo} were found to be
consistent with key characteristics of recent observations of
``Fermi-arc'' spectra in the pseudogap phase of the cuprates. In
particular, we showed that the peak in the spectral weight was
shifted away from the symmetrical $\vec{K}_p$ position. This shift
was linked to the same term in the Lagrangian that caused the shift
of the dispersion minimum in the VBS state, a term originally
introduced by Shraiman and Siggia \cite{ss}.

Finally, we note that our boundary CFT in Section~\ref{sec:qc} has
features that are reminiscent of self-consistent impurity models
\cite{qsi1} of `local' quantum phase transitions in Kondo lattice
systems. However, the details are different, and can only be
addressed in a quantum field-theoretic framework which we have
described. This work therefore also serves to place studies of
quantum transitions in the `dynamical mean field theory' approach
\cite{qsi1,gabi} in the field-theoretic context.

\acknowledgments

We are grateful to M.~Metlitski for valuable discussions. This
research was supported by the NSF grants DMR-0537077 (SS and RKK),
 DMR-0132874 (RKK) and DMR-0541988 (RKK). AK was
supported by Grant KO2335/1-1 under the Heisenberg Program of
Deutsche Forschungsgemeinschaft, ML by the Harvard Society of
Fellows, and TS by a DAE-SRC Outstanding Investigator Award in
India.


\appendix

\section{Single hole in the U(1) spin liquid}
\label{ribhu}

In this appendix, we calculate the dispersion of the holons due to
absorption and emission of the spinons. We will work within the
framework of the $t-J$ model Eq.~(\ref{tJmodel}), using the
formalism of Refs.~\onlinecite{klr,suyu}.

First consider the second term in Eq.~(\ref{tJmodel}), using a
Schwinger-boson mean-field theory~\cite{aa}, we arrive at
Eq.~(\ref{eq:schwinger}). After Fourier transformation, this
Hamiltonian can be diagonalized into the form
\begin{equation}H_{J}=\sum_{k\sigma}\epsilon_k
(\alpha^\dagger_{k\sigma}\alpha_{k\sigma} +
\beta^\dagger_{k\sigma}\beta_{k\sigma})\end{equation} by introducing
two new particles, \begin{eqnarray} \alpha_{k\sigma} &=& u_k
b_{k\sigma} + v_k \overline{b}^\dagger_{-k\sigma} \nonumber
\\ \beta_{k\sigma} &=& u_k \overline{b}_{k\sigma} + v_k
b^\dagger_{-k\sigma}\,, \end{eqnarray} where
\begin{equation} \gamma_k = \sum_\delta e^{ik\cdot \delta}=2
\left[\cos\left(\frac{k^\prime_x+ k^\prime_y}{2}a^\prime\right)
+\cos\left(\frac{k^\prime_x-k^\prime_y}{2}a^\prime\right)\right]
\end{equation} where $k^\prime_{x(y)}\in
[-\frac{\pi}{a^\prime},\frac{\pi}{a^\prime}]$ (i.e. BZ$^\prime$),
where $a^\prime = \sqrt{2}a$. $u(v)_k$ are real and
$u(v)_{k}=u(v)_{-k}$. A comparison of terms gives, $u_k=\cosh
\theta_k$ and $v_k=\sinh \theta_k$, where $\tanh 2\theta_k
=-Q\gamma_k/\lambda$. The dispersion of the $\alpha,\beta$
quasiparticles is $\epsilon_k = (\lambda^2-(Q\gamma_k)^2)^{1/2}$.
Using these we can calculate all the propagators of interest,
\begin{eqnarray}
-\langle b_{k\sigma} b^\dagger_{k\sigma}\rangle &=& -
\frac{i\nu_n+\lambda}{\nu_n^2+\epsilon_k^2}\equiv\mathcal{G}(k,i\nu_n)\\
-\langle b_{k\sigma} \overline{b}_{-k\sigma}\rangle &=&
-\frac{Q\gamma_k}{\nu_n^2+\epsilon_k^2}\equiv\mathcal{F}(k,i\nu_n)
\end{eqnarray}
We now define $\alpha=Q/\lambda$ and use this parameter instead of
$Q$. Then, {\em e.g.\/}, the dispersion becomes $\epsilon_k =
\lambda(1-(\alpha\gamma_k)^2)^{1/2}$, where $0<\alpha<0.25$.

\begin{figure}
  \includegraphics[width=3in]{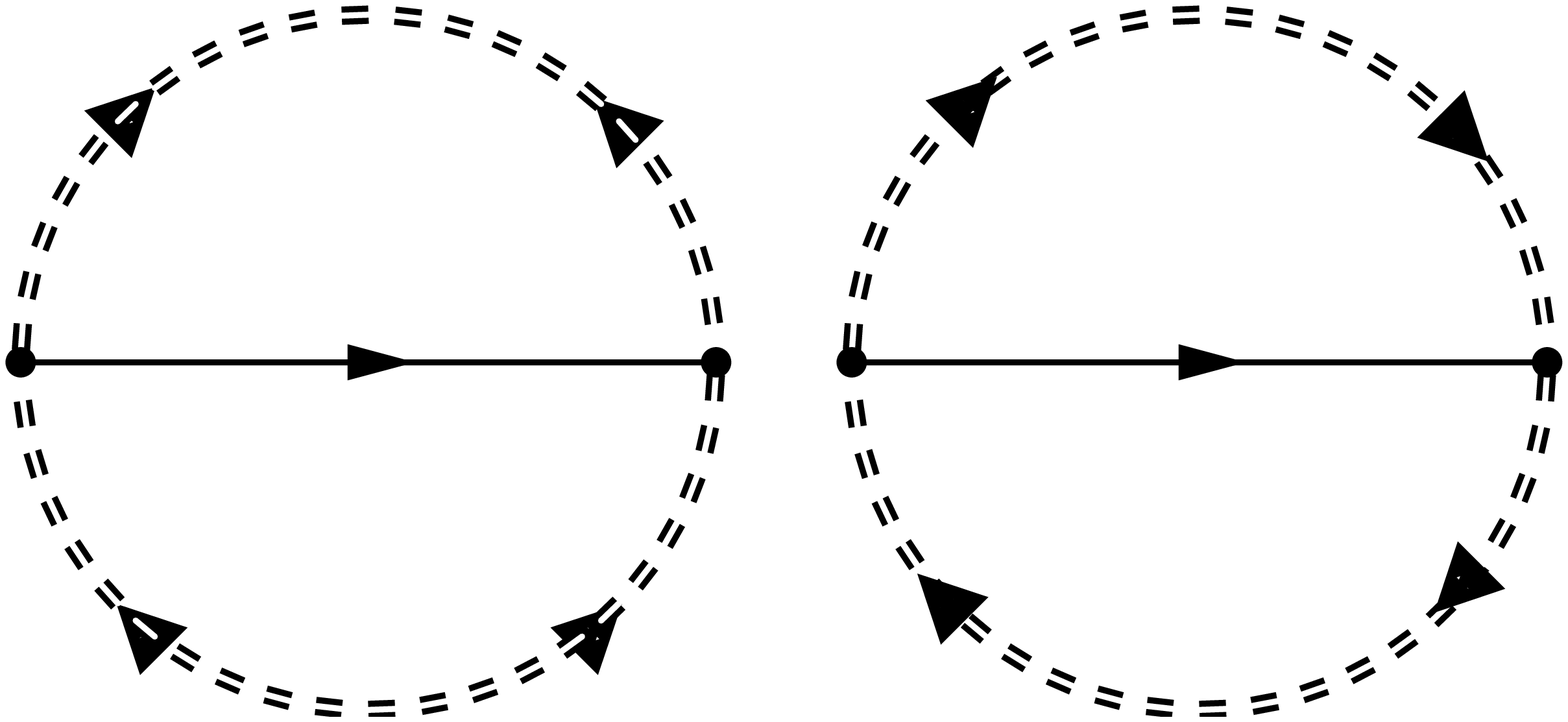}
\caption{The two diagrams contributing to the holon self-energy in
Eq.~(\ref{self-energy}). The solid lines are the holon propogators
and the double-dashed lines are the spinons.}
\label{fig:hlndsp_diag}
\end{figure}

The hopping term in the $t-J$ model in terms of Schwinger bosons
after Fourier transformation gives,
\begin{eqnarray}
H_{t}=t\sum_{k,q_1,q_2}\gamma_k
\varepsilon_{\sigma\sigma^\prime}(f_{+,q_1}b^\dagger_{k+q_1
\sigma}\overline{b}_{ k+q_2\sigma^\prime}f^\dagger_{-, q_2}+ f_{-,
q_1}\overline{b}^\dagger_{k+q_1 \sigma^\prime}b_{
k+q_2\sigma}f^\dagger_{+,q_2})
\end{eqnarray}

\begin{figure}
  \includegraphics[width=4in]{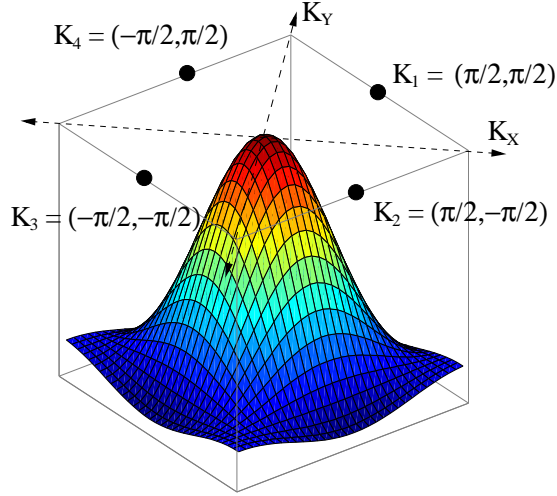}
\caption{(color online) The dispersion $\omega_k t^2/\lambda$
obtained from Eq.~(\ref{eq:simple_spec}). The points $K_p$ where the
holon dispersion have their minima are marked for clarity. We have
used $\alpha=0.24$ here, note however that these results are
generic.} \label{fig:spinw}
\end{figure}

Working to second order in the hopping, we now calculate the
self-energy of the $f_{\pm}$ particles due to the $H_t$. There are
two diagrams at order $t^2$ (shown in Fig.~\ref{fig:hlndsp_diag})
that need to be summed up, they evaluate to,
\begin{eqnarray}
\label{self-energy} \Sigma^A_1(k,i\omega_n)&=&[2]t^2
\frac{1}{\beta^2}\sum \gamma_{q-p}^2 \mathcal{G}(i\nu_m,q)\mathcal{G}(i\omega_n-i\omega_l+i\nu_m,k+q-p)\mathcal{G}^B_f(i\omega_l,p) \\
\Sigma^A_2(k,i\omega_n)&=&[-2]t^2 \frac{1}{\beta^2}\sum
\gamma_{q-p}\gamma_{q+k}
\mathcal{F}(i\nu_m,q)\mathcal{F}(i\omega_n-i\omega_l+i\nu_m,k+q-p)\mathcal{G}^B_f(i\omega_l,p)\nonumber
\end{eqnarray}
The factors in $[...]$ appear from the spin sums. The sum of the
above diagrams becomes (we assume here that $A(p,\omega<0)=0$),
\begin{equation}
\label{eq:seresult} \Sigma^A(i\omega_n,k)=
\sum_{q,p}[t^2(\gamma_{q-p}u_{k+q-p}v_q - \gamma_{q+k}u_q
v_{k+q-p})^2]\mathcal{G}^B_f(i\omega_n-\epsilon_q-\epsilon_{k+q-p},k)
\end{equation}
A full self-consistent solution of the above equation corresponds to
summing all diagrams that have non-crossing boson lines. The
dispersion of the f-particles is obtained by looking for the poles
of the interacting Greens' function, i.e. $\omega_k = \Re
[\Sigma(\omega_k,k)]$. We shall not undertake the fully
self-consistent calculation here, instead we simply use the self
energy to second-order in $t$ which gives.
\begin{equation}
\label{eq:simple_spec} \omega_k = \sum_{q,p} F_{t^2}(k,p,q)
\frac{1}{-\epsilon_{k+q-p}-\epsilon_q},
\end{equation}
where $F_{t^2}(k,p,q)= t^2(\gamma_{q-p}u_{k+q-p}v_q -
\gamma_{q+k}u_q v_{k+q-p})^2$. This result is plotted in
Fig.~\ref{fig:spinw}. As discussed in the text, the holon disperion
has minima at the $(\pm\pi/2,\pm\pi/2)$. The band masses at the
minima are highly anisotropic with the holon dispersion much lighter
along the zone diagonal than perpendicular to it.

\section{Critical properties of the $\mathbb{CP}^{N-1}$ model}
\label{cpn}

As argued elsewhere, and reviewed in Section~\ref{sec:intro}, the
critical theory of the N\'eel-VBS transition on the square lattice
is described by the $\mathbb{CP}^{N-1}$ model. Here we will review
the $1/N$ expansion of this theory, obtain an effective field theory
suitable for bound state problems, and compute a scaling dimension
in the boundary CFT associated with a charged impurity.

We are interested in the properties of the field theory
\begin{eqnarray}
Z &=& \int \mathcal{D} z_\alpha \mathcal{D} \lambda \mathcal{D}
A_\mu \exp \left( - \mathcal{S} \right) \nonumber \\
\mathcal{S} &=& \int d^2 r d \tau \left[ \frac{1}{g}
\left\{|(\partial_\mu - i A_\mu) z_\alpha |^2 + i \lambda (
|z_\alpha |^2 - N ) \right\}  \right] \label{cpaction}
\end{eqnarray}
Here the index $\alpha = 1 \ldots N$, and $\tau$ is imaginary time.
The theory has a conformally invariant critical point at a $g=g_c$,
and we want to describe the spectrum for $g>g_c$.

First, let us review the properties at $N=\infty$. The quantum
critical point is at $g=g_c$, where
\begin{equation}
\frac{1}{g_c} = \int_0^{\Lambda} \frac{d^3 p}{8 \pi^3} \frac{1}{p^2}
\end{equation}
Here, $\Lambda$ is an ultraviolet cutoff, and we will only be
interested in phenomena at scales much smaller than $\Lambda$.

For $g>g_c$, we have a large $N$ solution for the spin-gap U(1) spin
liquid with $\langle z_\alpha \rangle = 0$, $i \lambda =
\Delta_0^2$, where
\begin{equation}
\int_0^{\infty} \frac{d^3 p}{8 \pi^3} \left( \frac{1}{p^2} -
\frac{1}{p^2 + \Delta_0^2} \right) = \frac{1}{g_c} - \frac{1}{g}
\label{e4}
\end{equation}
Here $\Delta_0$ is a measure of the spin gap at $N=\infty$; we will
discuss the renormalization of this to the energy scale $\Delta$ in
the following subsection.

For $g<g_c$, we have the N\'eel phase, which breaks the SU($N$)
rotation symmetry with $\langle z_\alpha \rangle = m_0 \sqrt{N}
\delta_{\alpha,1}$ and $i \lambda = 0$. The value of $m_0$ is given
by
\begin{equation}
m_0^2 = \frac{1}{g} - \frac{1}{g_c}
\end{equation}
The properties of this phase are easy to understand by spin-wave
theory (``chiral perturbation theory''), and we won't consider it
further.

We now consider the structure of fluctuations for $g\geq g_c$. The
$1/N$ expansion can be easily generated by a Feynman graph expansion
of the theory
\begin{eqnarray}
\label{zcpn} Z &=& \int \mathcal{D} z_\alpha \mathcal{D} \lambda
\mathcal{D}
A_\mu \exp \left( - \mathcal{S} \right)  \\
\mathcal{S} &=& \int d^2 r d \tau \left[  \left\{|(\partial_\mu - i
A_\mu) z_\alpha |^2 + \Delta_0^2 |z_\alpha|^2 + i \lambda  |z_\alpha
|^2 \right\} \right] \nonumber \\
&~&~~~~~~~~~~~~+ \int \frac{d^3 p}{8 \pi^3} \left[ \frac{N}{2} \Pi
(p) |\lambda (p) |^2 + \frac{N}{2} K(p) A_\mu (-p) \left(
\delta_{\mu\nu} - \frac{p_\mu p_\nu}{p^2} \right) A_\nu (p) \right].
\nonumber
\end{eqnarray}
Here we have rescaled $z_\alpha$ by $\sqrt{g}$, and there is no
further dependence upon $g$. The kernels $\Pi(p)$ and $K(p)$
represent one-loop $z_\alpha$ contributions. We avoid double
counting in the Feynman graph expansion simply by avoiding any
further bubble self-energy contributions to the $\lambda$ and
$A_\mu$ propagators. The explicit values of the kernels are
\begin{eqnarray}
\Pi (p) &=& \int \frac{d^3 q}{8 \pi^3} \frac{1}{(q^2 +
\Delta_0^2)((p+q)^2 + \Delta_0^2)} \nonumber \\
&=& \frac{1}{4 \pi p} \cot^{-1} \left( \frac{2 \Delta_0}{p} \right)
\nonumber \\
&=& \left\{ \begin{array}{ccc} 1/(8 \pi \Delta_0) &~~& p \ll
\Delta_0 \\
1/(8p) &~~~& p \gg \Delta_0 \end{array}
 \right.
\end{eqnarray}
and
\begin{eqnarray}
K(p) &=& - 2 \int \frac{d^3 q}{8 \pi^3} \frac{q^2 - (p \cdot q)^2 /
p^2}{((q+p/2)^2 +
\Delta_0^2)((q-p/2)^2 + \Delta_0^2)} + 2 \int \frac{d^3 q}{8 \pi^3} \frac{1}{q^2 + \Delta_0^2} \nonumber \\
&=& \int_0^1 \frac{dx}{2 \pi} \left[ \sqrt{p^2 x(1-x) + \Delta_0^2}
-
\Delta_0 \right] \nonumber \\
&=& \frac{p^2 + 4 \Delta_0^2}{8 \pi p} \cot^{-1} \left( \frac{2
\Delta_0}{p} \right) - \frac{\Delta_0}{4 \pi}
\nonumber \\
&=& \left\{ \begin{array}{ccc} p^2/(24 \pi \Delta_0) &~~& p \ll
\Delta_0 \\
p/16 &~~~& p \gg \Delta_0 \end{array}
 \right. ,
\end{eqnarray}
where we have used the Feynman parameter method to evaluate the
integrals.

The following subsection will formulate an approach suitable for
bound state problems, and examine the response to an impurity.

\subsection{Non-relativistic effective field theory} \label{nrt}

We are interested here primarily in the physics at scales $p
\lesssim \Delta_0$. As in ordinary quantum electrodynamics, this
regime is conveniently expressed in terms of a non-relativistic
effective field theory. We will see that the low-velocity expansion
associated with the effective field theory is justified because the
typical velocity in a bound state scales as $1/\sqrt{N}$.

We first need to express the relativistic theory in (\ref{zcpn}) in
terms of the non-relativistic particle and anti-particle
eigenstates. To this end, we decouple the kinetic term for the
$z_\alpha$ by a Hubbard-Stratanovich field $\Pi_\alpha$:
\begin{equation}
|(\partial_\tau - i A_\tau) z_\alpha|^2 \rightarrow |\Pi_\alpha |^2
+ i \Pi_\alpha^\ast (\partial_\tau - i A_\tau) z_\alpha +
\mbox{c.c.}
\end{equation}
Then we parameterize
\begin{eqnarray}
z_\alpha &=& \frac{1}{\sqrt{2 \Delta_0}} \left( p_\alpha +
h_\alpha^\dagger \right) \nonumber \\
\Pi_\alpha &=& \frac{1}{\sqrt{2 \Delta_0}} \left( p_\alpha -
h_\alpha^\dagger \right) \, \label{nr1}
\end{eqnarray}
where $p_\alpha$ and $h^\alpha$ will be the spinon and anti-spinon
excitations. For the SU(2) case, with $N=2$, we will make the
replacement $h^\alpha \rightarrow \varepsilon^{\alpha \beta}
h_\beta$, so that both the spinons and anti-spinons transform under
the same representation of SU(2).

We now insert (\ref{nr1}) in the effective action (\ref{zcpn}), and
expand in powers of spatial gradients. The structure of this
expansion determines the form of the effective field theory. With
some additional approximations which will shortly be justified, we
postulate, up to order $1/N$, the following non-relativistic
effective field theory
\begin{eqnarray}
Z_{NR} &=& \int \mathcal{D} p_\alpha \mathcal{D} h_\alpha
\mathcal{D}
A_\tau \exp \left( - \mathcal{S}_{NR} \right) \nonumber \\
\mathcal{S}_{NR} &=& \int d^2 r d \tau \left[ \frac{N}{48 \pi
\Delta} (\vec{\nabla} A_\tau)^2  + p^{\alpha\dagger} \left(
\frac{\partial}{\partial \tau} - i A_\tau + \Delta \left(1+
\frac{\zeta_1}{N} \right) - \frac{1}{2 \Delta}
\vec{\nabla}^2 \right) p_\alpha \right. \nonumber \\
&~&~~~~~~~~~~~~~~+ \left. h^\dagger_\alpha \left(
\frac{\partial}{\partial \tau} + i A_\tau + \Delta \left(1+
\frac{\zeta_1}{N} \right) - \frac{1}{2 \Delta} \vec{\nabla}^2
\right) h^\alpha  \right]
\end{eqnarray}
Here we have {\em defined\/} the renormalized energy scale $\Delta$
so that the co-efficient of the electrostatic term $(\vec{\nabla}
A_\tau)^2$ term is $N/(48 \pi \Delta)$. At $N=\infty$, we have
$\Delta=\Delta_0$; there will be corrections at order $1/N$, which
can be computed from (\ref{zcpn}), which will depend upon the cutoff
$\Lambda$, and which ensure that $\Delta \sim (g-g_c)^{\nu}$, with
\cite{irkhin} $\nu = 1 - 48/(\pi^2 N)$. The numerical constant
$\zeta_1$ is a universal number to be determined by a matching
computation to the fully relativistic theory (\ref{zcpn}); we will
not carry out such a computation here.

The dominant interaction is the `Coulomb' interaction between the
$p_\alpha$ and the $h^\alpha$, mediated by the electrostatic
potential $A_\tau$. This leads to a potential energy between
opposite unit charges of the form
\begin{eqnarray}
V(r) &=& \frac{24 \pi \Delta}{N} \int^{~\sim \Delta} \frac{d^2 k}{4
\pi^2}
\frac{(1 - e^{i \vec{k} \cdot \vec{r}})}{k^2} \nonumber \\
&=& \frac{12 \Delta}{N} \left[ \ln \left( r \Delta \right) + \zeta_2
\right] \, . \label{vr}
\end{eqnarray}
Here $\zeta_2$ appears to be a second unknown constant, but we have
the freedom to set $\zeta_2=0$ without modifying any observable
property. This is because $Z_{NR}$ applies only to
spinon/anti-spinon configurations which have net charge zero, and
for these $\zeta_2$ can be absorbed into a redefinition of
$\zeta_1$.

 From this potential we see that the spinon and anti-spinon form a
bound state with a spatial extent $\sim \sqrt{N}/\Delta$. This means
that the velocity expansion holds in powers of $1/\sqrt{N}$. One can
also see from this that the coupling to the spatial components of
the gauge field, $\vec{A}$, lead to corrections which are higher
order in $1/N$ from the terms included. Short-range interaction
terms (induced from $\lambda$ fluctuations) like $|p_\alpha|^2
|h^\alpha|^2$ are also easily seen in perturbation theory to be
higher order: they induce corrections which depend upon the value of
the bound state function at the origin $\sim 1/\sqrt{N}$. These
arguments justify our dropping $\lambda$ and $\vec{A}$ fluctuations
in $Z_{NR}$.

\subsection{Boundary exponent of an impurity}
\label{qimp}

As motivated in Section~\ref{sec:qc}, we consider here the problem
of a static charged impurity coupled to the 2+1 dimensional critical
point of $\mathbb{CP}^{N-1}$ model. For this we modify the theory in
Eq.~(\ref{cpaction}) to
\begin{eqnarray}
\widetilde{Z} &=& \int \mathcal{D} z_\alpha \mathcal{D} f
\mathcal{D} \lambda \mathcal{D}
A_\mu \exp \left( - \widetilde{\mathcal{S}} \right) \nonumber \\
\label{GL} \widetilde{\mathcal{S}} &=&\int d^{2}r d\tau
\mathcal{L}_{z}+\int d \tau f^\dagger \left(
\frac{\partial}{\partial \tau} + \varepsilon_0 - i A_\tau (r=0,\tau)
\right) f,\\
\mathcal{L}_{z} &=&  |(\partial_\mu - i e_0 A_\mu) z_\alpha |^2 + s
|z_\alpha |^2 + \frac{u_0}{2} \left(|z_\alpha|^2
\right)^2  \nonumber \\
&~&~~~~~~~~~+ \frac{1}{2} \left( \epsilon_{\mu\nu\lambda}
\partial_\nu A_\lambda \right)^2 ,\nonumber
\end{eqnarray}
Note that the Grassman field $f$ depends only upon $\tau$, while the
bosonic fields vary over both $r$ and $\tau$. For completeness, we
add ``bare'' terms corresponding to the spinon interaction and the
gauge field kinetic energy, and we will see that the critical
exponents are independent of these terms. We will now compute the
scaling dimension $\eta_{h}/2$ of the composite operator
$h(\tau)=z_\alpha(r=0,\tau) f^\dagger(\tau)$.

At the critical point $s=s_{c}$, the bare boson propagator is
$G_{0}(k)=1/k^{2}$. The dressed photon propagator can be obtained by
summing over $z$-bubbles and has the form
\begin{equation}
\label{prop} D_{\mu\nu}(p)=(p^{2}+\gamma p)^{-1}\Big(\delta_{\mu\nu}
-\zeta\frac{p_{\mu}p_{\nu}}{p^{2}}\Big),
\end{equation}
where $\zeta$ is the arbitrary ``gauge parameter'' and
\[
\gamma=Ne_{0}^{2}/16.
\]
In a similar way, one can obtain the dressed $z$-field interaction
vertex
\[
V(k)=\frac{u_{0}}{1+2u_{0}N\Pi(k)},
\]
where $\Pi(k)=1/(8k)$ is the bosonic bubble. The large-$N$ expansion
is thus the expansion in dressed photon lines and dressed
$z$-vertices.

The anomalous dimensions $\eta_{z}$ and $\eta_{f}$ of the fields $z$
and $f$ can be read off as the coefficients in the contributions of
the type $-\eta_{z}p^{2}\ln(p)$ and $i\eta_{f}\omega \ln(\omega) $
to the self-energies $\Sigma_{z}(p)$ and $\Sigma_{f}(\omega)$ which
enter the corresponding dressed propagators
$G^{-1}(p)=p^{2}+\Sigma_{z}(p)$  and
$G_{f}^{-1}(\omega)=-i\omega+\Sigma_{f}(\omega)$ (here $\omega$ is
counted from the reference energy $\varepsilon_{0}$). To the order
of $1/N$, the self-energies are given by
\begin{eqnarray}
\label{sigma-z} \Sigma_{z}(p)&=&\int \frac{d^{3}k}{(2\pi)^{3}}
G_{0}(p-k)
\Big\{ 2V(k) \nonumber\\
&-&e_{0}^{2}(2p_{\mu}-k_{\mu})(2p_{\nu}-k_{\nu})D_{\mu\nu}(k)
\Big\},\\
\Sigma_{f}(\omega)&=&e_{0}^{2}\int
\frac{d^{3}k}{(2\pi)^{3}}\frac{1}{-i(\omega-\Omega)}
D_{00}\big(k=(q,\Omega)\big)
\end{eqnarray}
and after tedious but straightforward calculation one obtains
\begin{eqnarray}
\label{eta-z-f} &&
\eta_{z}=\frac{4}{3\pi^{2}N}-\frac{4e_{0}^{2}}{3\pi^{2}\gamma}
-(1-\zeta)\frac{e_{0}^{2}}{2\pi^{2}\gamma},\\
&& \eta_{f}=-(1+\zeta)\frac{e_{0}^{2}}{2\pi^{2}\gamma},
\end{eqnarray}
One should realize that, taken separately, $\eta_{z}$ and $\eta_{f}$
are both gauge-dependent and do not have any direct physical sense.
It turns out that there is no contribution to $\eta_{h}$ from the
vertex diagram describing an exchange of a photon between the $z$
and $f$ particles, so to the $1/N$ order, the scaling dimension
$\eta_h/2$ of $h$ is given by $1/2 + \eta_{z}+\eta_{f}$, which in
turn yields the (gauge-independent) expression (\ref{etah}) for the
boundary exponent $\eta_{h}$.


\end{document}